%% file: main.tex
\documentclass[aps,prd,twocolumn,amsfonts,amssymb,amsmath,showpacs,letterpaper,superscriptaddress,nofootinbib,altaffilletter,fontenc]{revtex4-2}
\usepackage[utf8]{inputenc}
\usepackage[T1]{fontenc}

\usepackage{graphicx}	
\usepackage{caption}
\usepackage{subcaption}
\usepackage{amsmath}	
\usepackage{amssymb}	
\usepackage{tabularx}
\usepackage{enumerate}
\usepackage{enumitem}
\usepackage{comment}
\usepackage{acronym}
\usepackage{physics}
\usepackage{multirow}
\usepackage{float}
\usepackage[citecolor=blue,linkcolor=blue,colorlinks=true,urlcolor=blue]{hyperref}
\usepackage[misc]{ifsym}
\usepackage{orcidlink}
\usepackage{tikz}
\usepackage{standalone}
\usepackage{xcolor}
\usepackage[normalem]{ulem}
\usetikzlibrary{arrows.meta,calc}

\definecolor{cbBlue}{HTML}{0072B2}
\definecolor{cbOrange}{HTML}{E69F00}
\definecolor{cbGreen}{HTML}{009E73}

\newacro{gw}[GW]{gravitational wave}
\newacro{imbh}[IMBH]{intermediate-mass black hole}
\newacro{smbh}[SMBH]{supermassive black hole}
\newacro{bhns}[BHNS]{black hole neutron star}
\newacro{bbh}[BBH]{binary black hole}
\newacro{bh}[BH]{black hole}
\newacro{bns}[BNS]{binary neutron star}
\acrodef{FAR}[FAR]{false alarm rate}
\newacro{bf}[BF]{Bayes' factor}
\newacro{cbc}[CBC]{compact binary coalescence}
\newacro{ce}[CE]{Cosmic Explorer}
\acrodef{SNe}[SNe]{Supernova}
\newacro{da}[DA]{data analysis}
\newacro{et}[ET]{Einstein Telescope}
\newacro{eob}[EOB]{Effective-One-Body}
\newacro{fd}[FD]{frequency domain}
\newacro{gw}[GW]{gravitational wave}
\newacro{gr}[GR]{General relativity}
\newacro{hm}[HM]{Higher mode}
\newacro{ifo}[IFO]{interferometer}
\newacro{imr}[IMR]{inspiral-merger-ringdown}
\newacro{im}[IM]{inspiral-to-merger}
\newacro{kagra}[KAGRA]{Kamioka Gravitational Wave Detector}
\newacro{ligo}[LIGO]{Laser Interferometer Gravitational-Wave Observatory}
\newacro{lso}[LSO]{Last Stable Orbit}
\newacro{lvc}[LVC]{LIGO-Virgo Collaboration}
\newacro{lvk}[LVK]{LIGO-Virgo-KAGRA}
\newacro{lo}[LO]{leading order}
\newacro{ns}[NS]{neutron star}
\newacro{nr}[NR]{numerical relativity}
\newacro{pn}[PN]{post-Newtonian}
\newacro{pe}[PE]{parameter estimation}
\newacro{psd}[PSD]{power spectral density}
\newacro{cwb}[cWB]{Coherent WaveBurst}
\newacro{far}[FAR]{false alarm rate}
\newacro{ifar}[iFAR]{inverse false alarm rate}
\newacro{ml}[ML]{machine learning}
\newacro{cnn}[CNN]{convolutional neural network}
\newacro{asd}[ASD]{amplitude spectral density}
\acrodef{KN}[KN]{kilonova}
\newacro{xg}[XG]{next-generation}
\newacro{jsd}[JSD]{jensen shannon divergence}
\newacro{qnm}[QNM]{quasi-normal mode}
\newacro{cwt}[CWT]{continuous wavelet transform}
\newacro{hlv}[HLV]{Hanford-Livingston-Virgo}
\newacro{ecmm}[ECMM]{effective chirp mass model}
\acrodefplural{KN}[KNe]{kilonovae}
\newacro{qc}[QC]{quasi-circular}
\newacro{snr}[SNR]{signal-to-noise ratio}
\acrodef{SNR}[SNR]{signal-to-noise ratio}
\newacro{ng}[NG]{Next Generation}
\newacro{eos}[EoS]{Equation of State}
\newacro{agn}[AGN]{Active Galactic Nuclei}

\definecolor{amethyst}{rgb}{0.6, 0.4, 0.8}

\definecolor{bittersweet}{rgb}{1.0, 0.44, 0.37}

\begin{document}

\title{Can model-agnostic signal reconstruction unmask eccentricity disguised as precession in IMBH mergers?}

\author{Shubhagata Bhaumik}
\email[E-mail: ]{shubhagata.bhaumik@iitb.ac.in}
\affiliation{Department of Physics, Indian Institute of Technology, Bombay, Powai, 400076, India}
\author{Johann Fernandes}
\email[E-mail: ]{johann.fernandes@iitb.ac.in}
\affiliation{Department of Physics, Indian Institute of Technology, Bombay, Powai, 400076, India}
\author{Archana Pai}
\email[E-mail: ]{archanap@iitb.ac.in}
\affiliation{Department of Physics, Indian Institute of Technology, Bombay, Powai, 400076, India}



\begin{abstract}
Orbital eccentricity in compact binary mergers serves as a smoking gun for dynamical formation. Massive \acp{bbh}, especially with remnant/component as an intermediate-mass \ac{bh} provide a suitable testbed for probing these signatures. 
One such candidate event was GW190521, whose merger produced a remnant \ac{imbh}. 
However, analysing high-mass eccentric \acp{bbh} in the ground-based interferometric detector network with quasi-circular precessing waveform models poses considerable challenges, as it can introduce severe biases in the inferred binary parameters, especially due to their short duration in the ground-based detector bands. In this work, we systematically investigate the eccentricity--precession degeneracy using a suite of non-spinning \ac{nr} simulations with GW190521-like source-frame total mass spanning three mass ratios, eccentricities $e_{\text{10Hz}} \le 0.32$ and inclinations $\iota \in \{0^\circ, 90^\circ\}$ under both real O4a noise ($\text{SNR} = 14.5$) and projected O5 noise ($\text{SNR} \approx 35$). To break this degeneracy, we use the unmodeled \ac{cwb} algorithm as a model-agnostic waveform consistency test. Additionally, we introduce a tail fraction statistic, $\eta_{0.05}$ derived from the \ac{cwb} waveform consistency test to quantify the resolvability of the mismatches between the reconstructed eccentric signal and quasi-circular, precessing \ac{pe} samples. The \ac{pe} results show that in realistic noise scenarios, eccentricity can be masked as spin precession, particularly for loud asymmetric systems with high eccentricity and prominent higher-order-mode content. We demonstrate that the waveform consistency test fails to  capture this bias and unmask eccentricity in O4 noise scenarios. In higher SNR scenarios, the test is highly effective in breaking this degeneracy, given the presence of higher-order mode content. 

\end{abstract}

\date[\relax]{Dated: \today }

\maketitle

\section{Introduction}

Since the first direct detection of a \ac{gw} in 2015 \cite{LIGOScientific:2016aoc, LIGOScientific:2016vlm}, the \ac{lvk} international scientific collaboration \cite{LIGOScientific:2014pky, VIRGO:2014yos, KAGRA:2020tym} has detected hundreds of \ac{gw} transient signals with data from successive runs of the network of advanced ground-based \ac{gw} detectors. With the recent release of the GWTC-5 catalog \cite{LIGOScientific:2026sit, LIGOScientific:2026ifv, LIGOScientific:2026wfs}, \ac{lvk} has now detected over 390 \ac{gw} signals \cite{LIGOScientific:2018mvr, LIGOScientific:2020ibl, KAGRA:2021vkt, LIGOScientific:2025slb, LIGOScientific:2026wfs}, the vast majority of which originate from \ac{bbh} coalescences \cite{LIGOScientific:2026sit}.

As the population of detected \ac{bbh} mergers grows, one of the key science goals is to determine their formation mechanism. Often, the detected \ac{gw} signals carry crucial clues hidden in the form of the astrophysical properties. For example, \acp{bbh} that evolve in isolated fields largely have individual black hole spins aligned to the orbital angular momentum \cite{Kalogera:1999tq, Cashen:2016neh, Farr:2017uvj}. In contrast, \ac{bbh} systems that form dynamically are expected to have an isotropic spin distribution due to random gravitational encounters \cite{Rodriguez:2016vmx, Doctor:2019ruh}. Similarly, component masses and mass-ratios can constrain our understanding of stellar evolution processes. For example, \acp{bbh} with a component mass in the pair-instability supernova (PISN) mass gap or exhibiting highly asymmetric mass ratios suggest hierarchical mergers in dense stellar environments or active galactic nucleus (AGN) disks \cite{2017ApJ...840L..24F, Kimball:2019mfs}.
By and large, orbital eccentricity imprints a confident and unique signature in the \ac{gw} signal emitted by a binary originating from a dynamical environment \cite{Rodriguez:2017pec, Tagawa:2020jnc, 2022Natur.603..237S}. Although, some studies have shown that an isolated channel can produce eccentric binaries via SN kicks \cite{1983ApJ...267..322H, Brandt_SN_kick}. While detecting evidences of orbital eccentricity can offer immense advantages in the inference of binary formation channels, incorporating eccentricity into waveform approximants has been challenging due to the increased dimensionality of the parameter space, more complex phase modulations, increased computational cost, and incomplete coverage in the broad spin-eccentricity parameter space.


The past decade has seen rapid advances in the development of analytical and numerical waveform models for \ac{gw} signals from \ac{bbh} mergers with orbital eccentricity. These includes the SEOBNR \citep{Khalil:2021txt, Ramos-Buades:2021adz, Gamboa:2024hli}, TEOBResumS \citep{Damour:2014sva, Chiaramello:2020ehz, Nagar:2021gss}, and IMRPhenom \citep{Planas:2025feq, Ramos-Buades:2026kbq} waveform families. Integrating these models into \ac{pe} frameworks has allowed for placing constraints on eccentricity for the catalog of detected events \cite{Iglesias:2022xfc, Romero-Shaw:2022xko, Gupte:2024jfe, Gupte:2026whi, Xu:2025ajj, Zeeshan:2026pga}. However, one key limitation that has remained is that the merger-ringdown regimes of these waveforms are currently calibrated with quasi-circular numerical relativity (NR) waveforms \cite{Carullo:2023kvj}. Due to this, these waveform models cannot be used to accurately constrain the eccentricity for short-duration signals originating from high-mass binary mergers, where the detectable waveform is dominated by the merger and ringdown rather than the inspiral. This is especially the case for binaries formed in dense dynamical environments that retain non-negligible eccentricity right before merger, as several studies have shown that the merger amplitude depends strongly on the residual eccentricity \citep{Carullo:2023kvj, Bhaumik:2024cec}. 

In case of inferring the source properties for such short-duration signals, many physical hypotheses can fit the signal well, which can lead to a large number of degenerate astrophysical interpretations and incorrect downstream analyses, like population inference. This fundamental challenge is best illustrated by the high-mass \ac{bbh} merger event GW190521 and the extensive follow-up studies dedicated to its interpretation.

\subsection{GW190521 Interpretations}

GW190521 was the most massive \ac{bbh} detected during the first three observing runs by the \ac{lvk} collaboration \cite{LIGOScientific:2020iuh, LIGOScientific:2020ufj}. The candidate was detected on May 19th, 2019, during the third observing run (O3) of advanced LIGO and Virgo detectors. It was also the first direct detection of an \ac{imbh} system, with its remnant \ac{bh} as an \ac{imbh}. The detected signal was found to be consistent with source-frame component masses of $85^{+21}_{-14}~ M_{\odot}$ and $66^{+16}_{-17}~ M_{\odot}$ (total mass, $M_{\text{tot}} \sim 150\,M_{\odot}$), remnant mass $142^{+28}_{-16}~ M_{\odot}$,  effective aligned-spin parameter, $\chi_{\text{eff}} = 0.08^{+0.27}_{-0.36}$, and effective precession-spin parameter, $\chi_{\text{p}} = 0.68^{+0.25}_{-0.37}$.
Due to its short in-band duration, various physical interpretations have been proposed to explain the origin of this event. In this section, we review key results and interpretations proposed in the literature.
\begin{enumerate}
    \item \textbf{Orbital Eccentricity:} Several studies propose that GW190521 originates from an eccentric \ac{bbh} merger. For example, \citet{Gayathri:2020coq} performed Bayesian inference using eccentric \ac{nr} simulations, concluding that GW190521 was most consistent with a highly eccentric ($e=0.69$) merger, with source-frame component masses $78^{+9}_{-5}~ M_{\odot}$ and $78^{+9}_{-5}~ M_{\odot}$ and spin parameters $\chi_{\text{eff}}=0.27^{+0.56}_{-0.51}$ and $\chi_{\text{p}} = 0.66^{+0.28}_{-0.60}$. Following a similar approach with a recently extended \ac{nr} simulations catalog, the eccentricity constraint was obtained to be $e_{\text{20Hz}} = 0.207^{+0.192}_{-0.147}$ by \citet{McMillin:2026qiq}.
    \citet{Romero-Shaw:2020thy} performed Bayesian inference with a semi-analytical aligned-spin eccentric waveform model (SEOBNRE), finding that the data prefers eccentricity $e \geq 0.1$ to a quasi-circular hypothesis. \ac{pe} with this waveform yields source-frame component masses of $92^{+26}_{-16}~ M_{\odot}$ and $19^{+18}_{-19}~ M_{\odot}$ and an effective spin $\chi_{\text{eff}}=0.0^{+0.2}_{-0.2}$. Several analyses with waveforms that contain higher-order modes (\texttt{TEOBResumS-Dali} and \texttt{SEOBNRv5EHM}) have however concluded that the GW190521 signal does not prefer an eccentric hypothesis over a quasi-circular one \citep{Iglesias:2022xfc, Gupte:2024jfe, Pompili:2026yxq}. Although, more recently, \citet{Jan:2026mnp} demonstrated using a suite of eccentric waveform approximants (\texttt{TEOBResumS-Dali}, \texttt{IMRPhenomTEHM}, and \texttt{SEOBNRv5EHM}) that with a wider eccentricity prior, the eccentric hypothesis is preferred, with $e_{\text{5Hz}}=0.81^{+0.05}_{-0.01}$. They additionally show that the previous non-eccentric interpretations originated from restrictive eccentricity prior bounds ($e_{\text{5Hz}} \le 0.4$).   
    
    \item \textbf{Spin Precession:} Spin misalignment relative to the orbital angular momentum offers an alternative explanation for GW190521. Parameter estimation performed by LVK Collaboration yielded an informative $\chi_{\text{p}}$ posterior \citep{LIGOScientific:2020iuh}. \citet{Miller:2023ncs} performed inference on data truncated at varying cut-off times around coalescence to pinpoint which part of the signal constrains the precession measurement. They demonstrated that neither pre-merger nor post-merger data alone constrains $\chi_{\text{p}}$, and that the constraint originates from contrasting the quiet final pre-merger cycle with the loud merger--ringdown.
    
    \item \textbf{Hyperbolic Encounter:} \citet{Gamba:2021gap} analyzed GW190521 under a hyperbolic-encounter hypothesis, finding that the data is consistent with two BHs with source-frame component masses  $85^{+88}_{-22}~ M_{\odot}$ and $59^{+18}_{-37}~ M_{\odot}$ undergoing two encounters before merging into an IMBH. Subsequently, \citet{Lange:2026eqx} reanalysed GW190521 using an extended version of the model that includes higher-order modes, demonstrating that the signal continues to favour a dynamical-capture scenario over a quasi-circular spin-precessing one.

    \item \textbf{Intermediate Mass-Ratio Inspiral:} \citet{Nitz:2020mga} analysed GW190521 considering two different priors for the binary's masses: 1) uniform in mass ratio and source-frame total mass, and 2) uniform in source-frame component masses. They obtain source-frame component masses of $168^{+15}_{-61}~ M_{\odot}$ and $16^{+33}_{-3}~ M_{\odot}$ with the first choice of mass priors. The second choice of priors yields a bimodal mass posterior with source-frame component masses of the low mass-ratio mode as $100^{+17}_{-18}~ M_{\odot}$ and $57^{+17}_{-16}~ M_{\odot}$ and the high mass-ratio mode as $166^{+16}_{-35}~ M_{\odot}$ and $16^{+14}_{-3}~ M_{\odot}$. Their studies find a bimodal posterior for the mass ratio for the uniform in component masses prior, with the max-$\mathcal{L}$ parameters in the  high-mass ratio ($q \geq 4$) mode.
    \citet{Chandra:2023nge} performed a follow-up analysis using simulated signals from \ac{imbh} binary systems in ground-based detectors and demonstrated that the choice of priors can introduce substantial biases in the inferred total mass, mass ratio and luminosity distance. This study emphasises the importance of analysing such events with a diverse set of priors to prevent biases in the detector-frame mass posteriors.
 
    \item \textbf{Exotic Compact Objects:} Going beyond the \ac{bbh} hypothesis, \citet{CalderonBustillo:2020fyi} find that GW190521 is most consistent with a head-on collision between two horizonless vector boson stars with detector-frame masses of $117^{+7}_{-8}~ M_{\odot}$ and $111^{+7}_{-15}~ M_{\odot}$, and a final spin of $0.75^{0.08}_{-0.04}$.
\end{enumerate}

Motivated by these diverse interpretations of GW190521, this work investigates the degeneracy between orbital eccentricity and spin-precession in the \ac{gw} signals emitted from IMBH binary systems \cite{CalderonBustillo:2020xms, Romero-Shaw:2022fbf, Xu:2025ajj}. 
To systematically investigate this degeneracy, we utilize a set of state-of-the-art eccentric NR simulations \cite{Scheel:2025jct, Hinder:2017sxy, SXS_Collab_2019} covering a grid of mass ratios ($q \in \{1.0, 0.5, 0.33\}$) and eccentricities ($e \in [0.02, 0.32]$) evaluated at both face-on ($\iota = 0^\circ$) and edge-on ($\iota = 90^\circ$) orientations to inject signals in realistic detector noise. We perform Bayesian inference on these injections using a quasi-spherical waveform model to characterize how unmodeled eccentricity can falsely show up as constraints in spin precession.

To break this degeneracy, we employ the \ac{cwb} algorithm as a model-agnostic waveform consistency test \cite{Salemi:2019uea, Szczepanczyk:2020osv}. By comparing the \ac{cwb} reconstructions of our target injections with the \ac{pe} posterior samples, we evaluate whether this test is effective in breaking the eccentricity-spin-precession degeneracy for the injection cases considered.

The paper is organized as follows. Section~\ref{sec:methods} details our simulation study framework, quasi-spherical Bayesian inference setup, the \ac{cwb} waveform consistency test, and the $\eta_{0.05}$ tail statistic. In section~\ref{sec:results}, we present the results of our parameter estimation studies, largely focusing on the biases in the inferred spin precession. In section~\ref{sec:cwb_reconstruction_results}, we analyze the results of our waveform consistency test using the on-source and off-source \ac{cwb} overlap distributions and finally, we demonstrate the utility of the $\eta_{0.05}$ tail fraction statistic. In section~\ref{sec:conclusions}, we discuss the implications of our findings in the context of current and future detector scenarios, and summarize our conclusions.

\section{Methodology}
\label{sec:methods}
We employ a suite of gravitational-wave data-analysis techniques in this work. In the following subsections, we briefly describe these methods.

\subsection{Bayesian Inference}
\label{sec:methods_pe}

Given a detected \ac{gw} transient signal, we perform an inference study to estimate the astrophysical parameters with an existing waveform model. In \ac{gw} astrophysics, we perform Bayesian \ac{pe} which is a straightforward application of Bayes' theorem. 

Given some data $D$ and a model $M$ characterised by parameters $\Theta$, one can obtain the probability of the data being described by parameters $\Theta$ under $M$. This distribution is referred to as the posterior which we denote by $p(\Theta|D,M)$. We use Bayes' theorem to express the posterior as

\begin{equation}
p(\Theta|D,M) = \frac{\mathcal{L}(D|\Theta,M)\ \pi(\Theta|M)}{Z} \label{eq:bayes_theorem} 
\end{equation}

where $\mathcal{L}(D|\Theta,M)$ is the underlying data generation distribution conditioned upon $\Theta$ and is called as the \textit{likelihood}. $\pi(\Theta,M)$ is the prior and takes into account our prior beliefs about the distribution of the parameters $\Theta$. Finally, the evidence $Z=\int\mathcal{L}(D|\Theta,M)\pi(\Theta,M)\dd\theta$ is responsible for ensuring that $p(\theta|D,M)$ is a genuine probability distribution by normalising the numerator.

Constructing the likelihood is not a trivial exercise as it requires modelling the data generation process. The simplest approximation is to model the noise in the detector as a zero-mean wide-sense stationary coloured Gaussian process which naturally leads to the Whittle likelihood \cite{whittle1951hypothesis, Veitch:2014wba}. In \ac{gw} \ac{pe}, the data $D$ is a sampled timeseries ($\bf d$) and is modelled as the sum of a noise ($\bf n$) and signal ($\bf s$) timeseries. The signal itself just serves to shift the mean of the data generation process. However, given that we do not know the signal a priori, we instead replace ($\bf s$) with a waveform model ($\bf h$) parametrized by $\theta$ in the sum.

\begin{equation}
{\bf d} = {\bf n} + {\bf h}    
\end{equation}

\begin{equation}
\begin{split}
\mathcal{L}(\mathbf{d} \mid \theta, S_n(f), M) = \exp \sum_i \biggl[ & \frac{-2|\tilde{\mathbf{d}}_i - \tilde{\mathbf{h}}_i|^2}{T S_n(f_i)} \\
& - \frac{1}{2}\log\left(\frac{\pi T S_n(f_i)}{2}\right) \biggr]
\end{split}
\end{equation}

\noindent
where tildes denote the discrete fourier transform counterparts of the timeseries. $S_n(f_i)$ is the \ac{psd} estimated from the data at the frequency bin $f_i$ and $T$ is the total analysis duration. One typically assumes that the \ac{psd} is known and has been estimated from a noise-only stretch of data rendering the second term a constant that is absorbed into the likelihood normalisation \cite{Veitch:2014wba}. This gives us our final likelihood expression for a known \ac{psd}. Dropping the explicit model dependence for brevity we have

\begin{equation}
\mathcal{L}({\bf d}|\theta, S_n(f)) = \exp\sum_i\left[\frac{-2|{\bf \tilde d}_i-{\bf \tilde h}_i|^2}{T S_n(f_i)}\right] \label{eq:likelihood}    
\end{equation}

Depending on the assumptions, \ac{cbc} \ac{gw} models are parameterised in 15 dimensions for quasi-spherical \footnote{quasi-circular precessing} systems. If eccentricity is included, the number of dimensions jumps to 17 \footnote{Eccentricity and mean anomaly}. This makes the computation of the posterior analytically intractable. Numerical methods such as Markov Chain Monte Carlo, Nested Sampling, Dynamic Nested Sampling, etc are thus employed to generate samples from the posterior\cite{emcee, nestedsamplingjohn, Higson:2018cwj}. The final output of a parameter estimation exercise is thus a collection of  points sampled from the posterior.

In this work, we investigate the impact of performing quasi-spherical parameter estimation on \ac{gw} signals originating from eccentric \acp{bbh}. We use Bilby with the dynesty sampler for sampling our posteriors \cite{bilby_paper, bilby_doi, speagle2020dynesty}. Throughout this work, we choose fixed priors over the whole 15 dimensional parameter space analogous to the choices made in the analysis of GW190521 albeit spanning a larger area in the mass prior space. In particular, we use uniform detector frame component mass priors as $\mathcal{U}[5,300]M_\odot$ with constraints on both mass ratio and detector frame total mass as $q\in [1/6,1]$ and $M_{T,\mathrm{det}}\in[80,400]M_\odot$, respectively. We choose luminosity distance priors corresponding to a uniform merger rate density in the source-comoving frame ranging from 50 to 10000 Mpc, to ensure coverage over our injection distribution. For the other parameters, we stick to the default precessing priors specified by Bilby. These settings ensure we stay consistent with \ac{lvk} \ac{pe} settings and that the results we obtain here can be readily extrapolated to a production analysis done at scale \cite{KAGRA:2021vkt, LIGOScientific:2025zdk, LIGOScientific:2026ifv}. For the sake of completeness, we summarise in Table \ref{tab:priors} all the chosen priors used in this study. 

\onecolumngrid
\vspace{\baselineskip}
\noindent
\begin{minipage}{\textwidth}
\begin{tabular*}{\textwidth}{@{\extracolsep{\fill}}lccc}
\hline
Parameter Name & Symbol & Prior & Constraint \\
\hline
Detector frame masses [$M_\odot$] & $m_{1,\mathrm{det}},m_{2,\mathrm{det}}$ & $\mathcal{U}[5,300]$ & $M_{T,\mathrm{det}}\in [80,400]\,M_\odot,\ q\in [0.1,1]$ \\
Spin magnitudes & $a_1,a_2$ & $\mathcal{U}[0,0.99]$ & - \\
Spin tilt angles [$\mathrm{rad}$] & $\theta_1,\theta_2$ & $\mathrm{Sin}[0,\pi]$ & - \\
Spin azimuthal angles [$\mathrm{rad}$] & $\phi_{12},\phi_{JL}$ & $\mathcal{U}[0,2\pi]$ & - \\
Ra, Dec [$\mathrm{rad}$] & $\alpha, \delta$ & Isotropic & - \\
Inclination angle [$\mathrm{rad}$] & $\theta_{JN}$ & $\mathrm{Sin}[0,\pi]$ & - \\
Polarization angle [$\mathrm{rad}$] & $\psi$ & $\mathcal{U}[0,\pi]$ & - \\
Coalescence phase [$\mathrm{rad}$] & $\phi_c$ & $\mathcal{U}[0,2\pi]$ & - \\
Luminosity distance [$\mathrm{Mpc}$] & $d_L$ & $\mathrm{UniformSourceFrame}[50,10000]$ & - \\
Geocentric Coalescence time [$\mathrm{s}$] & $t_c$ & $\mathcal{U}[t_0-0.1,t_0+0.1]$ & - \\
\hline
\end{tabular*}
\captionof{table}{Summary of the prior distributions and constraints used for the \ac{pe} \label{tab:priors}}
\end{minipage}
\vspace{\baselineskip}
\twocolumngrid

\subsection{cWB Reconstruction}
\label{sec:methods_cwb}

\subsubsection{Coherent WaveBurst}
\label{sec:methods_cwb_algorithm}

\ac{cwb} \cite{Klimenko:2005xv, 2008CQGra..25k4029K,cWB_MaxL, 2016PhRvD..93d3007T, Klimenko:2022nji} is a model-agnostic search algorithm that detects \ac{gw} signals by searching for coherent excess power in the time-frequency representation of data across the network of detectors. Since it does not rely on waveform templates, \ac{cwb} has been the only pipeline used in \ac{lvk} observing runs to conduct dedicated searches for eccentric compact binary coalescences \cite{LIGOScientific:2019dag, LIGOScientific:2023lpe}. \ac{cwb} first transforms whitened multi-detector strain data into a multi-resolution time-frequency representation using the Wilson--Daubechies--Meyer wavelet transform~\citep{Necula:2012zz} and clusters excess-energy pixels that are coherent across detectors. For each detected candidate, the algorithm maximizes a constrained likelihood ratio over sky locations and wave polarization states, and then applies an inverse wavelet transform to reconstruct the point-estimate time-domain \ac{gw} signal in each detector. Because this reconstruction requires no assumptions of the \ac{gw} signal's source parameters, it has previously been leveraged to probe missing physics such as orbital eccentricity \cite{Gayathri:2020coq, Das:2024zib}, higher-order modes in \ac{bbh} signals \cite{Halim:2021wgw}, and to probe waveform consistency across approximants for detected signals \cite{Salemi:2019uea, Szczepanczyk:2020osv}. Note that other deep-learning based frameworks have also recently been developed for reconstructing \ac{gw} signals with precession and higher order modes \cite{Chatterjee:2024alf, Chatterjee:2024mdv}.
In this work, we similarly leverage this capability of \ac{cwb} to check for consistency between a model-agnostic \ac{cwb} reconstruction of the target eccentric signal and the posterior samples inferred by \ac{pe}. We employ the version of \ac{cwb} that was used in the first part of the fourth observing run (O4a) \cite{LIGOScientific:2025slb}.

\subsubsection{The Waveform Consistency Test}
\label{sec:methods_cwb_waveform_consistency}

As described in Section~\ref{sec:methods_pe}, for each injected signal in our simulation set, we perform Bayesian \ac{pe} with the quasi-circular, precessing waveform model \texttt{IMRPhenomXPHM} \citep{Pratten:2020ceb}, which gives us posterior samples. We draw a random subset of $N$ from these samples, $\{\theta_i\}_{i=1}^{N}$, and inject the corresponding waveforms $h(\theta_i)$ into \textit{off-source} data, taken from a 14-day window (\textit{chunk}) around the true injection in which the \ac{psd} variation is negligible. We use the \ac{cwb} algorithm to reconstruct each of these off-source injections, giving $\hat{h}_i(t)$. The off-source reconstructions enable us to characterize systematics of \ac{cwb}-reconstruction errors as well as noise fluctuations.

For each off-source injection, we compare the whitened \ac{cwb}-reconstructed waveform $\hat{h}_i(t)$ with the whitened waveform $h(\theta_i; t)$ of the $i$-th posterior sample across the detector network. The consistency between two whitened waveforms $a$ and $b$ is quantified by the network overlap
\begin{equation}
    \mathcal{O}\!\left[a, b\right] = \frac{\langle a | b \rangle}{\sqrt{\langle a | a \rangle \langle b | b \rangle}},
\end{equation}
where the scalar product $\langle a | b \rangle$ is defined in the time-domain as
\begin{equation}
    \langle a | b \rangle = \sum_{j} \int_{t_1}^{t_2} a_{j}(t)\, b_{j}(t) \dd t .
\end{equation} 
where $a_j$ and $b_j$ are the whitened waveforms in detector $j$, and $[t_1,t_2]$ corresponds to the cWB reconstruction window.

 
The off-source overlaps
\begin{equation}
    \mathcal{O}_{\rm off}^{(i)} = \mathcal{O}\!\left[h(\theta_i), \hat{h}_i\right]
\end{equation}
form an empirical null distribution, $P(\mathcal{O}_{\text{off}})$, which accounts for \ac{cwb} reconstruction systematics and non-Gaussian detector noise.

We use this null distribution to quantify signal consistency by comparing it against the \textit{on-source} distribution $P(\mathcal{O}_{\text{on}})$, obtained by computing the overlaps between the \ac{cwb} reconstruction of the injected signal, $\hat{h}_s$, and the same \ac{pe} samples:
\begin{equation}
    \mathcal{O}_{\rm on}^{(i)} = \mathcal{O}\!\left[h(\theta_i), \hat{h}_s\right].
\end{equation}

\begin{figure}[htbp]
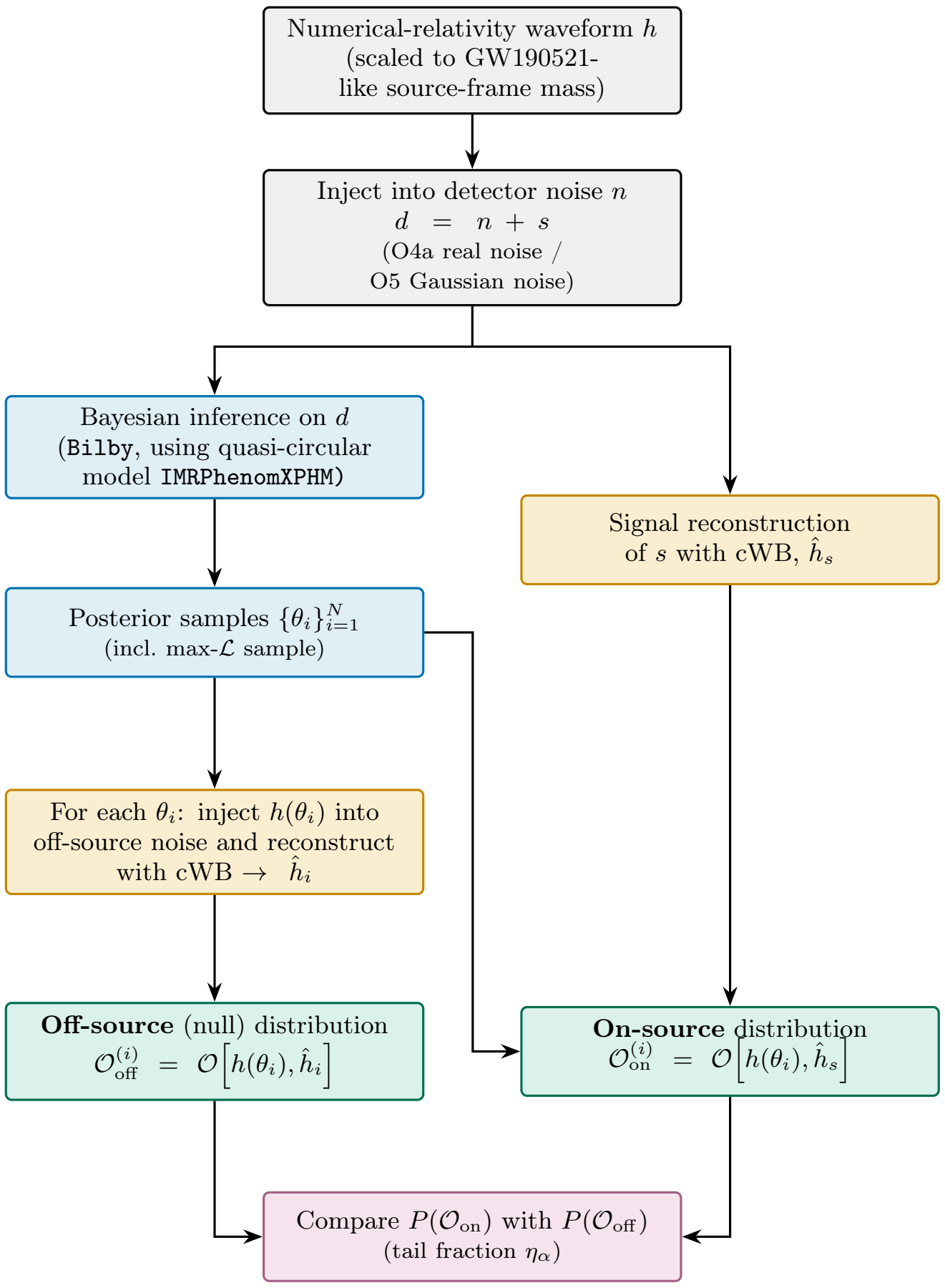

  \centering
  \includestandalone[width=\linewidth]{flowchart}
  \caption{Overview of the end-to-end workflow}
  \label{fig:flowchart}
\end{figure}

We explain the workflow for each injected signal using the flow chart in Figure~\ref{fig:flowchart}.

We can evaluate a single scalar $p$-value statistic that represents the percentile of the off-source (null) distribution $\mathcal{O}_{\text{off}}$ lying below the overlap between the target event's reconstruction and its $\text{max-}\mathcal{L}$ PE template $\mathcal{O}_{\text{on}}^{(\text{max-}\mathcal{L})}$:
\begin{equation}
    p = \frac{1}{N_{\text{off}}} \sum_{i=1}^{N_{\text{off}}} \Theta\left( \mathcal{O}_{\text{on}}^{(\text{max-}\mathcal{L})} - \mathcal{O}_{\text{off}}^{(i)} \right)
\end{equation}
where $\Theta(\cdot)$ is the Heaviside step function, and $N_{\text{off}}$ is the total number of samples comprising the off-source distribution. \ac{gw} catalogs so far have typically reported this single scalar $p$-value \cite{Salemi:2019uea, LIGOScientific:2020ibl, KAGRA:2021vkt, LIGOScientific:2025slb, LIGOScientific:2026ifv}. 

However, representing the on-source distribution with a single point estimate ignores parameter-estimation uncertainties. To mitigate this, we introduce the overlap \textit{tail fraction statistic} $\eta_{\alpha}$: 
\begin{equation} \eta_{\alpha} = \frac{1}{N_{\text{on}}} \sum_{i=1}^{N_{\text{off}}} \Theta\left( \mathcal{O}_{\text{off}}^{(\alpha)} - \mathcal{O}_{\text{on}}^{(i)} \right) \end{equation} 

\noindent
where $\Theta(\cdot)$ is the Heaviside step function, and $\mathcal{O}_{\text{off}}^{(\alpha)}$ denotes the $\alpha$-quantile of the off-source null distribution (we set $\alpha = 0.05$, corresponding to the 5th percentile). It measures the fraction of the posterior samples whose overlap with the reconstruction falls in the lowest $\alpha$ tail of what \ac{cwb} achieves for PE sample waveforms.

\subsection{Injection Set}
\label{sec:methods_injection_set}

To evaluate the impact of higher-order mode content and understand the effect of signal recovery across different observational regimes, we inject these waveforms under two inclination angle scenarios; {\it face-on ($\iota=0^\circ$) and edge-on ($\iota=90^\circ$)}.
While the face-on configuration is dominated by the $(\ell, m) = (2, \pm 2)$ quadrupole mode, the edge-on configuration significantly enhances the relative contribution of the higher-order modes ($\ell = 3, 4$ and $m = 0$).

\subsubsection{Quasi-circular, Precessing Injection (Control injection)}
\label{sec:methods_control_injection}

To establish a baseline reference, we perform the exact analyses described in the sections~\ref{sec:methods_pe} and \ref{sec:methods_cwb} for a quasi-spherical ($e = 0$) control injection. For this purpose, we select the \ac{nr} simulation \texttt{SXS:BBH:0391} ($q = 0.5$, $e = 0$, $\chi_{\text{p}}=0.34$ at $f_{\text{low}}=11 \, \text{Hz}$). This control set up serves two main purposes:

\begin{enumerate}
    \item {\it Null Calibration for the Waveform Consistency Test:} For a non-eccentric signal, the \ac{pe} samples and the \ac{cwb} reconstruction should ideally be in strong agreement. Consequently, we expect that the on-source overlap distribution $\mathcal{O}_{\text{on}}$ will be well-aligned with the off-source null distribution ($\mathcal{O}_{\text{off}}$), yielding a low tail fraction statistic, $\eta_{0.05}$. For reference, identical distributions would give $\eta_{0.05}=0.05$. 
    
    \item {\it Isolate Waveform Systematics:} This becomes especially relevant in high ac{snr} scenarios, where analyzing an $e = 0$ NR signal with a quasi-circular approximant (in our case, the \texttt{IMRPhenomXPHM} waveform approximant) isolates intrinsic approximant limitations and higher-order mode mismatches.
\end{enumerate}

\subsubsection{Eccentric Injections}
\label{sec:methods_eccentric_injections}

We consider nine non-spinning NR eBBH waveforms from the 3rd SXS Catalog \cite{Scheel:2025jct} to generate eccentric signals for injections. Due to the fact that these simulations contain multiple cycles, we were able to employ a standardized definition of eccentricity that was computed using the \texttt{gw-eccentricity} package \cite{Shaikh:2023ypz, Shaikh:2025tae}. The eccentricities quoted throughout this paper are at $f_{\mathrm{low}} = 10 \, \text{Hz}$ unless stated otherwise. The chosen simulation set covers eccentricities ($0 \leq e < 0.32$) and mass ratios ($1.0, 0.5, 0.33$). Additional details of each of these NR simulations are listed in Appendix~\ref{sec:appendix_nr_set}.

The injected signals contain higher-order spherical harmonic modes up to $\ell = 4$, explicitly including the $m=0$ modes, which are available in the most recent SXS catalog update. The $m = 0$ modes contain the non-oscillatory \ac{gw} memory, whose amplitudes and morphologies depend strongly on the orbital eccentricity, mass ratio, and binary inclination angle $\iota$.
All remaining intrinsic and extrinsic parameters of the injected signals, such as the source-frame total mass ($M_{\text{src}} = 150\,M_\odot$), sky location, and polarization angle are fixed to max-$\mathcal {L}$ parameters inferred from the LVK analysis of GW190521 \cite{LIGOScientific:2020iuh}. 

\subsection{Noise Scenarios}
\label{sec:methods_noise_scenarios}
We consider the following current and future detector scenarios.
\begin{enumerate}
\item{\textit{O4a Real Noise Baseline} (\ac{snr}$ = 14.5$):}
We choose a clean data segment of O4a data, free from non-stationary noise artefacts, or \textit{glitches} and inject signals at a fixed GPS time of 1379824500. To model current observing conditions and evaluate the recovery of a GW190521-like signal, we inject the \ac{gw} signals in real O4a detector data. For every mass ratio $q$, eccentricity $e$, and inclination $\iota$ combination, we adjust the luminosity distance $d_L$ (and consequently the redshift, $z$), to maintain a constant network GW190521-like signal-to-noise ratio of $\text{SNR} = 14.5$. The detector-frame total mass is adjusted accordingly as $M_{\text{det}} = M_{\text{src}}(1+z)$, where $M_{\text{src}}$ is fixed for all injections to a GW190521-like source-frame total mass of $M_{\text{src}} = 150\,M_\odot$. 

\item \textit{O5 Gaussian Noise Projection} ($\mathrm{SNR} \approx 35$): To project enhanced sensitivity during future observing runs, we consider the same injection configurations and distances in simulated, projected O5 Gaussian noise \cite{o5_psd_white_paper}. Due to the upgraded detector sensitivity curves, the resulting network signal-to-noise ratio scales up to $\text{SNR} \approx 35$, enabling us to evaluate how increased SNR and (marginally) longer durations impact parameter inference and \ac{cwb} signal reconstruction.
\end{enumerate}

We emphasize that the single noise-realization injections performed in this study are designed to model realistic single-event detection scenarios rather than to establish scaling laws with eccentricity. Because random noise fluctuations in a single realization can alter the posteriors, variations across eccentricity values should not be interpreted as universal trends. Isolating trends as a function of eccentricity or mass ratio alone would require zero-noise injections. Our primary goal, however, is to evaluate the joint impact of signal morphology and background noise in realistic observational contexts. 


\section{Results}
\label{sec:results}

\subsection{Control Injection}
\label{sec:results_control_injection}

\subsubsection{PE Results}
\label{sec:results_control_injection_pe}

As a baseline, we first examine the \ac{pe} results for the control injection (non-eccentric, precessing simulation).

\begin{figure}[htb!]
    \centering
    \begin{subfigure}[b]{0.48\columnwidth}
        \centering
        \includegraphics[width=\linewidth]{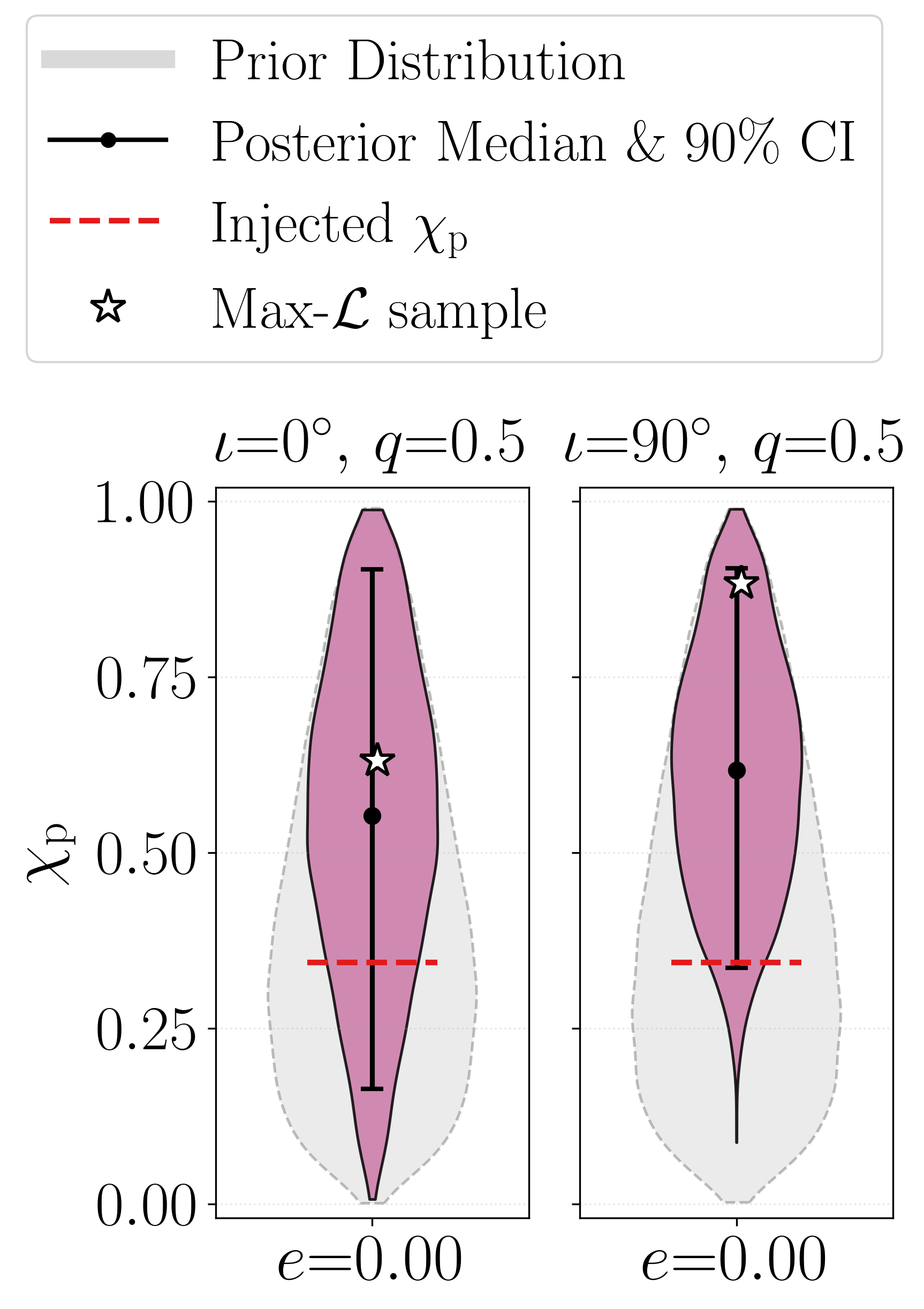}
        \caption{O4 Scenario}
        \label{fig:pe_control_chi_p_o4}
    \end{subfigure}
    \hfill
    \begin{subfigure}[b]{0.48\columnwidth}
        \centering
        \includegraphics[width=\linewidth]{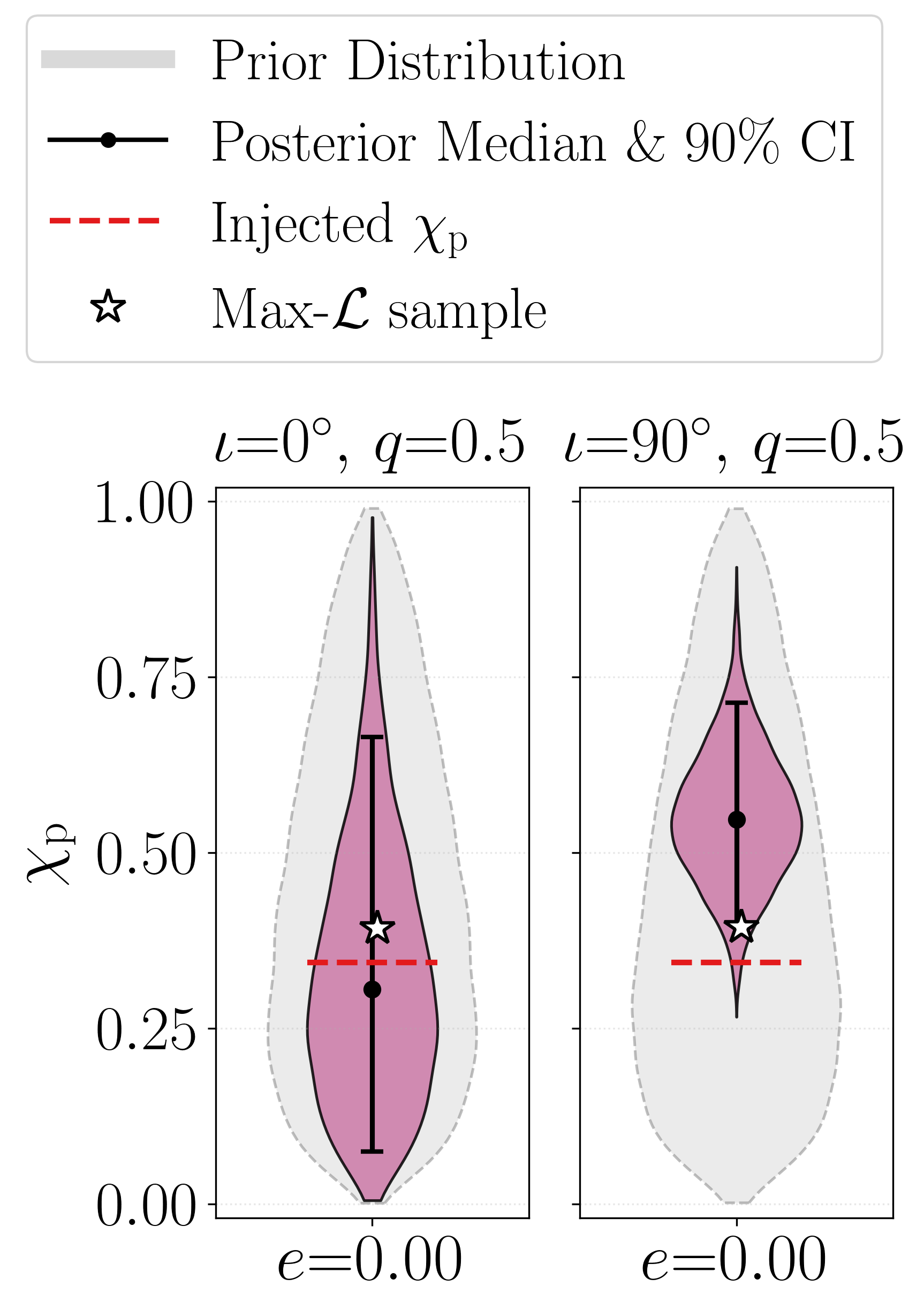}
        \caption{O5 Scenario}
        \label{fig:pe_control_chi_p_o5}
    \end{subfigure}
    \caption{Posterior probability distributions for the spin precession parameter ($\chi_{\text{p}}$) under the (a) O4 noise scenario ($\mathrm{SNR} = 14.5$) and (b) O5 noise scenario ($\mathrm{SNR} \approx 35$). Posteriors are shown as violin plots for both inclination configurations as indicated in the title. The dashed gray curve represents the $\chi_{\mathrm{p}}$ prior. For both O4 and O5 noise scenarios, in the face-on ($\iota=0^\circ$) configuration, the injected value lies inside the 90\% C.I. In the edge-on ($\iota=90^\circ$) configuration, the injected value lies outside the 90\% C.I. in the O5 noise scenario.}
    \label{fig:pe_control_chi_p_o4_o5}
\end{figure}

Figure~\ref{fig:pe_control_chi_p_o4_o5} shows the $\chi_{\text{p}}$ posteriors recovered for the face-on ($\iota=0^\circ$) and edge-on ($\iota=90^\circ$) configurations in both the O4 and O5 noise. 
For face-on systems, the injected $\chi_{\text{p}}$ lies within the 90\% credible interval for both noise scenarios. In the edge-on case, the injected value lies marginally inside the 90\% credible interval in the O4 case. In the O5 noise, where the
posterior is also narrower due to the higher \ac{snr}, the injected value lies outside the 90\% credible interval. We investigate possible systematics origins in Appendix~\ref{sec:appendix_pe_systematics}. Across both configurations and noise scenarios, the posterior differs from the prior, indicating that the data are informative about $\chi_{\text{p}}$ even though the recovery may not be exact (edge-on inclination, O5 noise case).

\subsubsection{Waveform Consistency Results}
\label{sec:results_control_injection_cwb}

\begin{figure}[h]
    \centering
    \begin{subfigure}[b]{0.48\columnwidth}
        \centering
        \includegraphics[width=\linewidth]{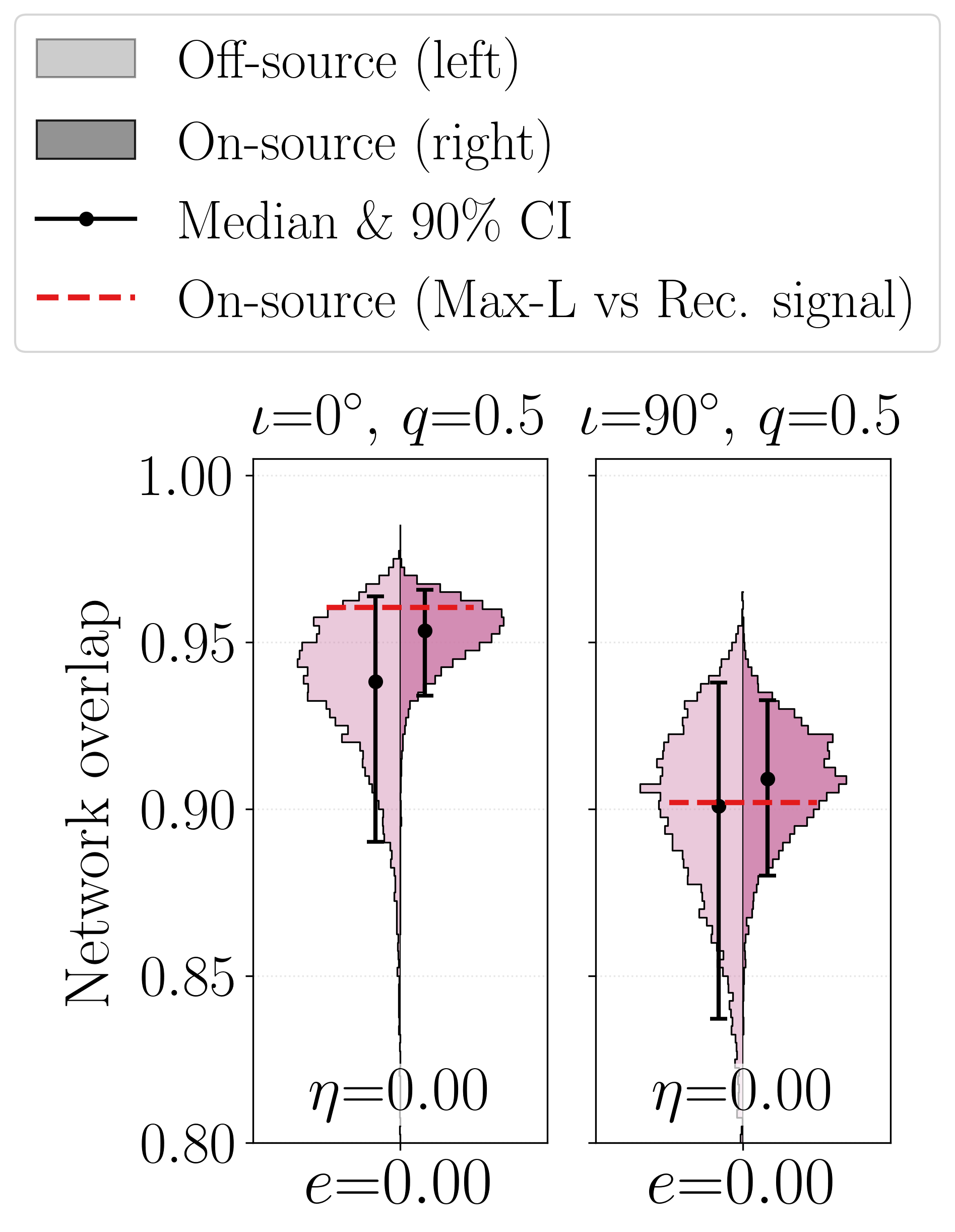}
        \caption{O4 Scenario}
        \label{fig:overlaps_violin_control_o5}
    \end{subfigure}
    \hfill
    \begin{subfigure}[b]{0.48\columnwidth}
        \centering
        \includegraphics[width=\linewidth]{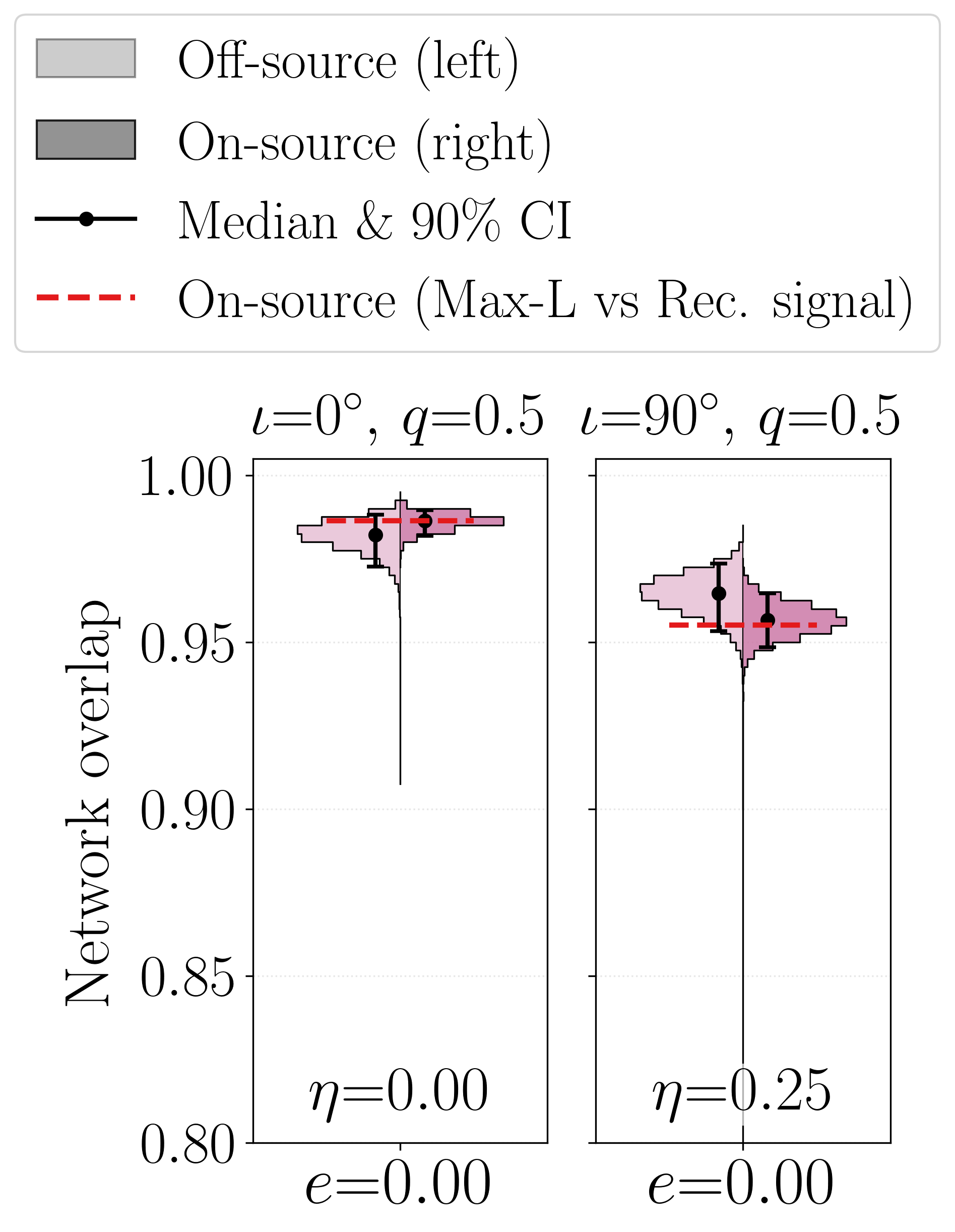}
        \caption{O5 Scenario}
        \label{fig:overlaps_violin_control_o4}
    \end{subfigure}
    \caption{Off-source and on-source overlap distributions for the control injection with the face-on configuration and edge-on configuration in the O4 and O5 noise scenarios shown in panels (a) and (b) respectively. With the exception of the edge-on configuration in the O5 noise scenario where $\eta_{0.05}=0.25$, $\eta_{0.05}=0.0$ for all cases.}
    \label{fig:overlaps_violin_control_o4_o5}
\end{figure}

We now examine the waveform consistency test results on the control injection. Figure \ref{fig:overlaps_violin_control_o4_o5} compares the on-source overlap distribution $P(\mathcal{O}_{\text{on}})$ with the off-source null
$P(\mathcal{O}_{\text{off}})$ for the face-on and edge-on configurations in both O4 and O5 noise. In all four cases, the two distributions are largely well-aligned. This is reflected in the overlap tail fraction $\eta_{0.05}$, which is $\eta_{0.05}=0.0$ throughout, with the exception of the edge-on configuration in O5 noise, where $\eta_{0.05}=0.25$. In the O5 scenario, since the distributions become sharper owing to the higher SNR, the consistency test is able to resolve subtle mismatches between the quasi-circular, precessing \ac{pe} templates and the reconstructed signal. A biased $\chi_{\text{p}}$ therefore does not by itself lead to a large inconsistency in the overlap distributions: the quasi-circular precessing \ac{pe} still reproduces the signal as reconstructed by cWB. This is in line with the expectation that large values of $\eta_{0.05}$ (section~\ref{sec:cwb_reconstruction_results}) should occur in the presence unmodeled eccentricity and not a generic \ac{pe} bias.

\subsection{Precession \& Eccentricity Mimicry}
\label{sec:pe_spin_results}

We now examine the recovered $\chi_{\text{p}}$ posteriors for the eccentric, non-spinning injection suite described in section~\ref{sec:methods_injection_set} for both noise scenarios and inclination configurations.

\subsubsection{O4 Scenario}
\label{sec:pe_spin_results_o4}
\begin{figure}[ht!]
\centering
\captionsetup[subfigure]{skip=5pt}  
    \begin{subfigure}{0.48\textwidth}
        \centering
        \includegraphics[width=\linewidth]{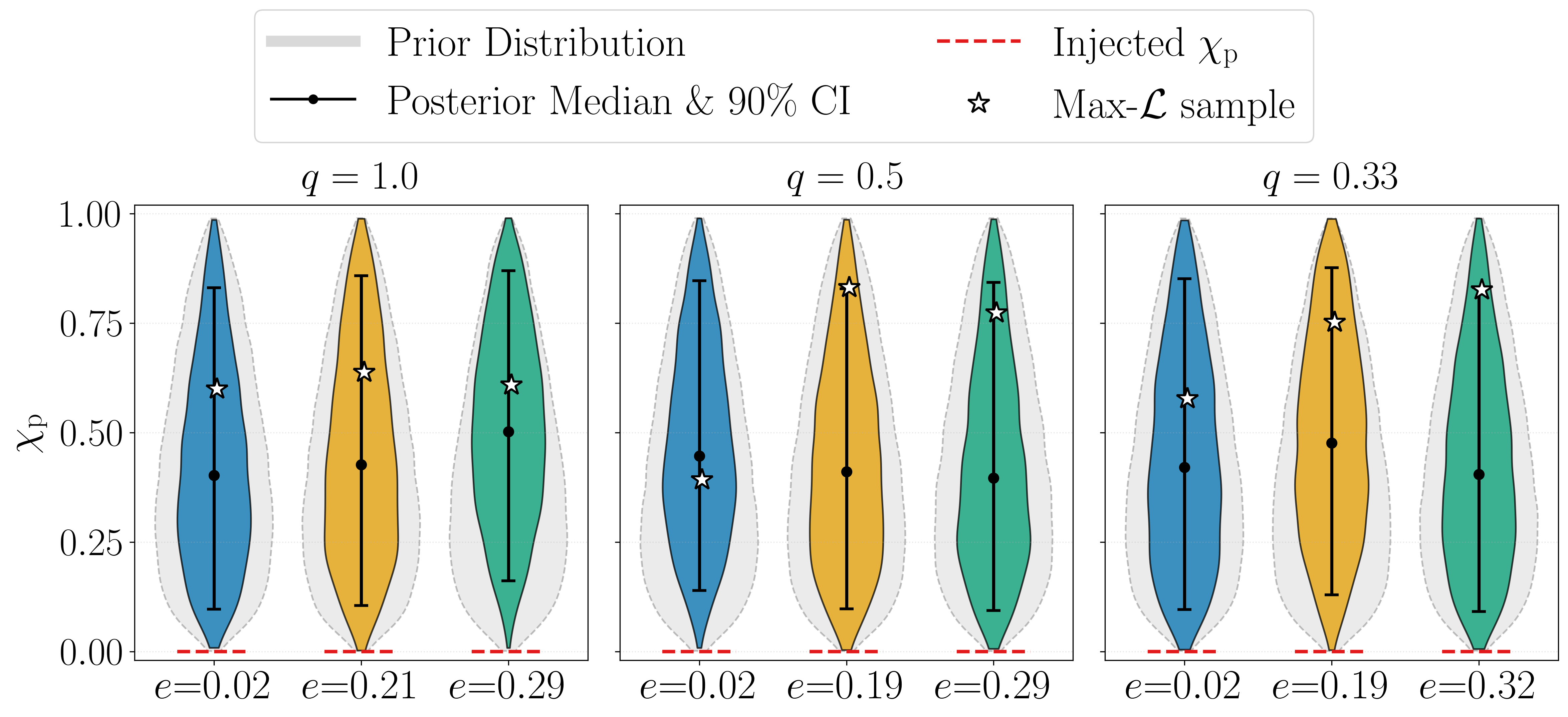}
        \caption{Face-on ($\iota = 0^\circ$) configuration}
        \label{fig:chi_p_posterior_o4_iota0}
    \end{subfigure}
    \vspace{0.1cm}
    \begin{subfigure}{0.48\textwidth}
        \centering
        \includegraphics[width=\linewidth]{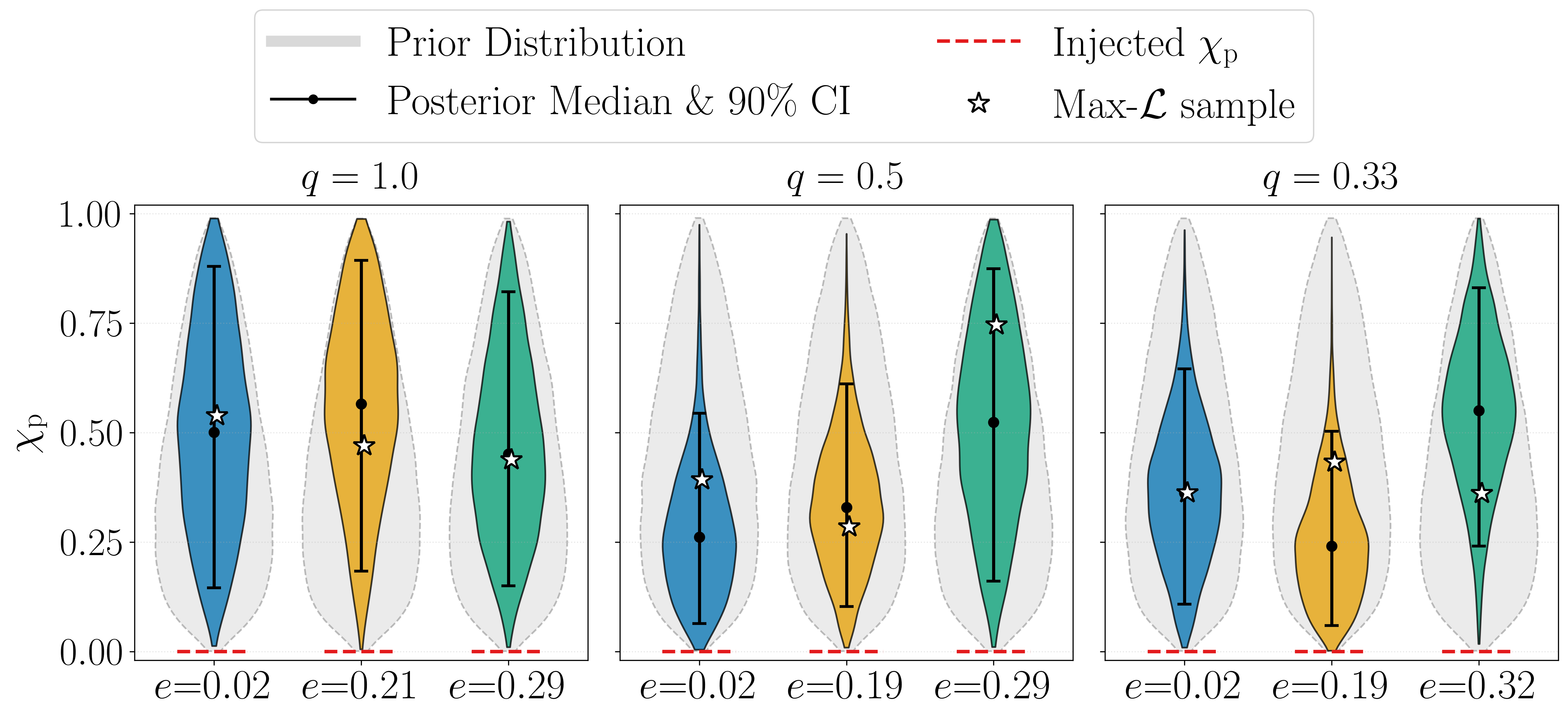}
        \caption{Edge-on ($\iota = 90^\circ$) configuration}
        \label{fig:chi_p_posterior_o4_iota90}
    \end{subfigure} 
    \caption{Posterior probability distributions for the spin precession parameter ($\chi_{\mathrm{p}}$) under the O4 noise scenario ($\mathrm{SNR} \approx 14.5$). Posteriors are shown as violin plots across mass ratios and eccentricities. The dashed gray curve represents the $\chi_{\mathrm{p}}$ prior. In panel (a), posteriors remain uninformative and track the prior across all nine configurations. In contrast, in panel (b), asymmetric mass ratios ($q \le 0.5$) show clear departures from the prior as unmodeled eccentric features in the higher-order modes are absorbed into the precessing degrees of freedom.}
    \label{fig:chi_p_posterior_o4}
\end{figure}

For all nine cases with different mass ratios and eccentricities, the recovered posteriors for $\chi_{\text{p}}$ are shown in Figure \ref{fig:chi_p_posterior_o4}. The gray dashed curves represent the $\chi_{\text{p}}$ priors resulting from the isotropic spin priors used.
For face-on configurations ($\iota=0^\circ$), the recovered $\chi_{\text{p}}$ posteriors are essentially uninformative, i.e., across all mass-ratios and eccentricities, the posterior distributions for the $\iota=0^\circ$ case are nearly identical to the prior, indicating that the data provides minimal constraint on spin precession when the signal is dominated by the $(\ell, m) = (2, \pm 2)$ mode. 

In contrast, for edge-on configurations ($\iota=90^\circ$), the $\chi_{\text{p}}$ posteriors display deviations from the prior distribution, particularly for the unequal mass scenarios ($q=0.5, 0.33$). This informativeness arises from increased higher-order mode content. 
The direction of the shift in the posterior from the prior depends on the injected eccentricity. For low and moderate eccentricities ($e \approx 0.02, 0.2$), the $\chi_{\text{p}}$ posteriors in general peak at lower values than the prior, effectively ruling out strong precession. At higher eccentricities, however, the $\chi_{\text{p}}$ posteriors peak at a higher value compared to the prior, hinting at spurious $\chi_{\text{p}}$ constraints.
Because the quasi-spherical PE model lacks explicit eccentricity parameters, it can misinterpret these higher-mode eccentric features as evidence of spin precession. 
Thus, for the O4 injection scenario, the recovery of $\chi_{\text{p}}$ demonstrates a strong dependence on the inclination angle $\iota$, or equivalently, the higher-order mode content which is captured by the $\chi_{\text{p}}$ parameter.


\subsubsection{O5 Scenario}
Under the upgraded sensitivity of the O5 observing run ($\mathrm{SNR} \approx 35$), the increased SNR significantly affects the recovery of spin precession, $\chi_{\mathrm{p}}$.

Unlike the O4 scenario in which face-on binaries ($\iota = 0^\circ$) yielded entirely uninformative posteriors, the O5 scenario exhibits noticeable deviations in the $\chi_{\mathrm{p}}$ posterior from the prior, even for the face-on configuration. At high SNR, higher-harmonic features and phase modulations induced by orbital eccentricity become more observable. Consequently, for face-on configurations with moderate-to-high eccentricities ($e \gtrsim 0.20$) and asymmetric mass ratios ($q = 0.5, 0.33$), the recovered $\chi_{\mathrm{p}}$ posteriors begin to deviate from the prior, indicating that a signal with high SNR alone can force a precessing quasi-circular model to fit eccentric features using false spin precession.

For edge-on configurations ($\iota = 90^\circ$), this departure from the prior becomes more pronounced.
Similar to the O4 scenario, the direction of the shift in the posterior from the prior depends on the injected eccentricity. In the O5 scenario, we see the $\chi_{\text{p}}$ posteriors shift to higher values than the prior for both moderate and high eccentricities ($e\approx 0.2, 0.3$). For the low eccentricity cases, the $\chi_{\text{p}}$ posteriors peak at lower values than the prior, effectively ruling out strong precession.
The combination of high SNR ($\mathrm{SNR} \approx 35$) and higher-order modes ($\ell = 3,4$ and $m = 0$) leads to higher-harmonic and non-oscillatory features in the data which are induced by orbital eccentricity. Because \texttt{IMRPhenomXPHM} lacks eccentricity parameters to capture these features, the sampler heavily leverages the precessing degrees of freedom. This results in tightly constrained, informative $\chi_{\mathrm{p}}$ posteriors, specifically for the moderate and high eccentric ($e \approx 0.2, 0.3$) cases, demonstrating that \textit{precession–eccentricity mimicry becomes more severe with increasing detector sensitivity and binary inclination.}


\begin{figure}[ht!]
\centering
\captionsetup[subfigure]{skip=5pt}  
    \begin{subfigure}{0.49\textwidth}
        \centering
        \includegraphics[width=\linewidth]{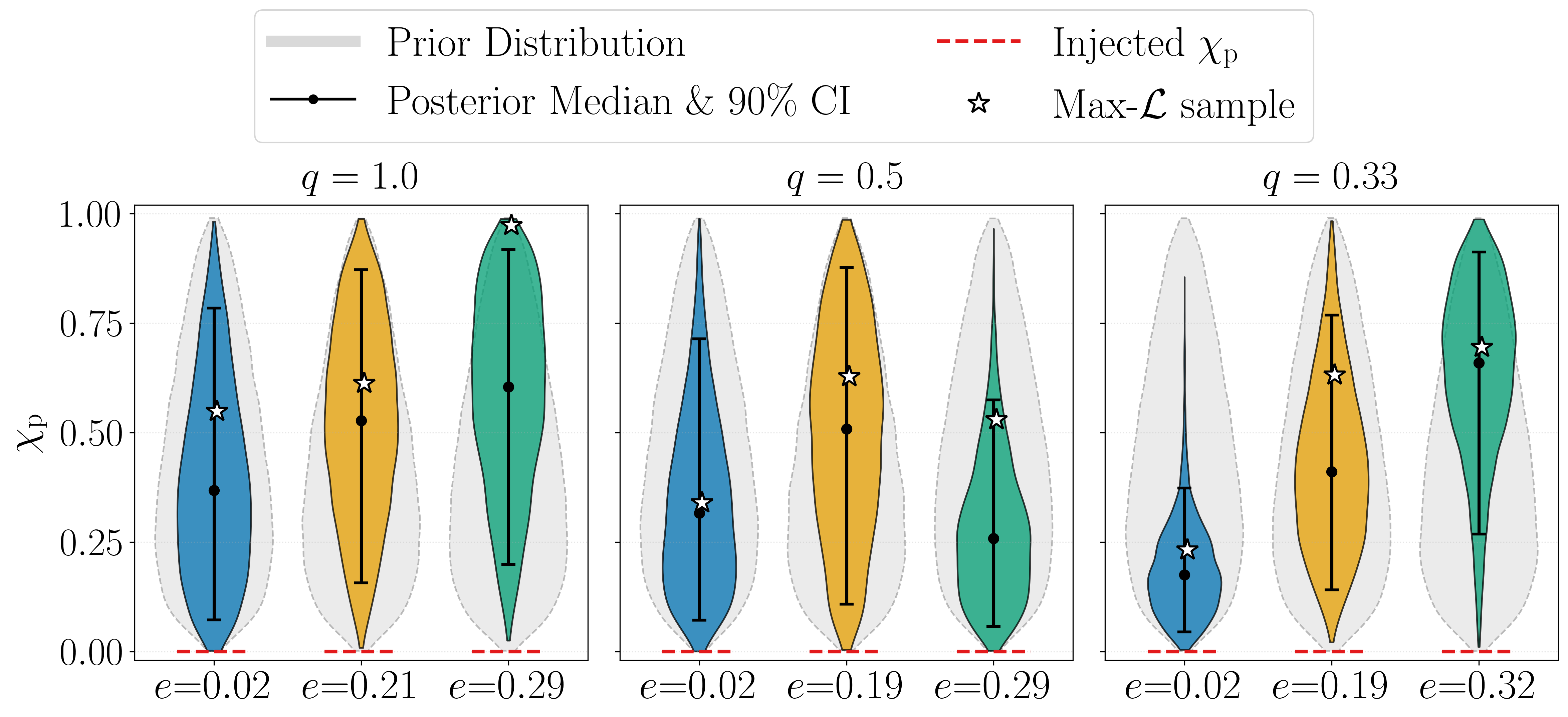}
        \caption{Face-on ($\iota = 0^\circ$) configuration}
        \label{fig:chi_p_posterior_o5_iota0}
    \end{subfigure}
    \vspace{0.1cm}
    \begin{subfigure}{0.49\textwidth}
        \centering
        \includegraphics[width=\linewidth]{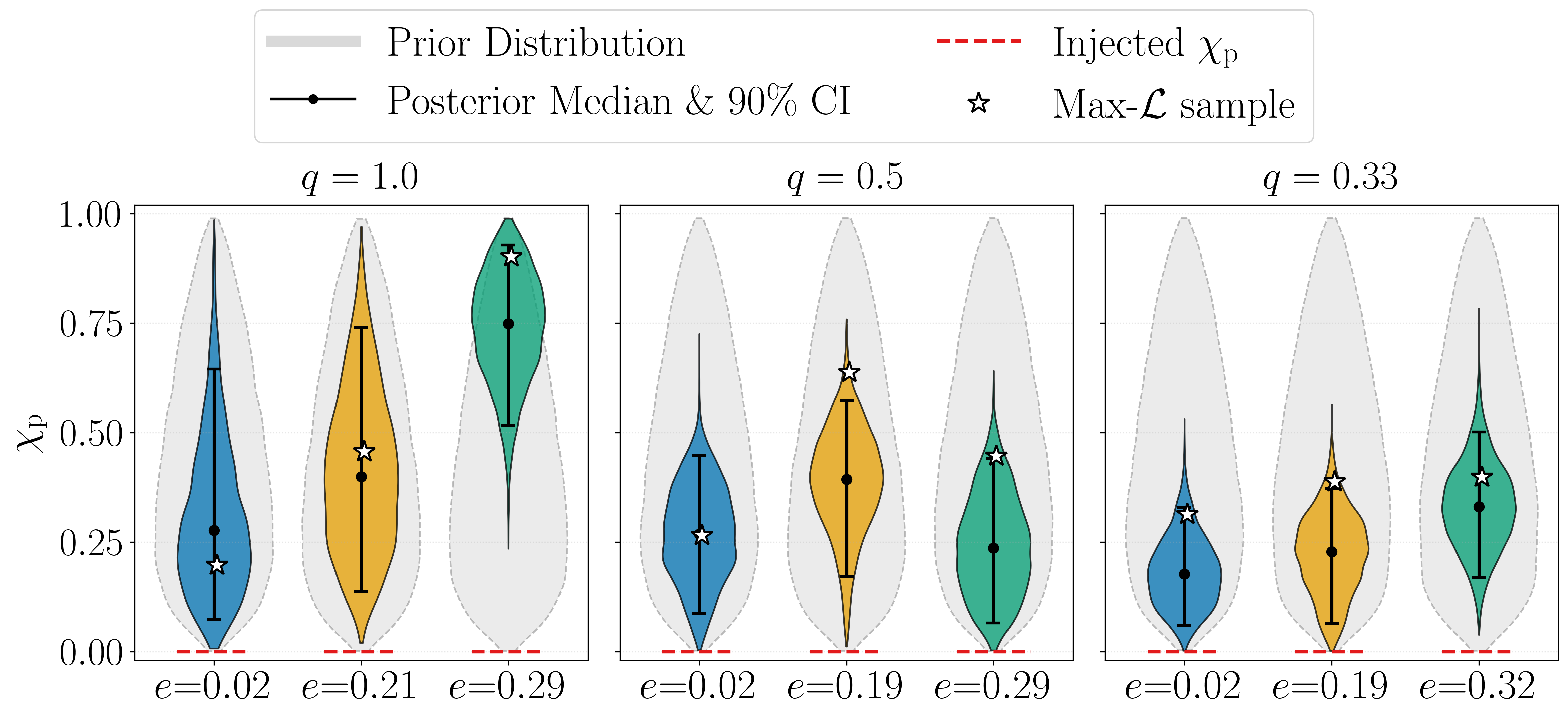}
        \caption{Edge-on ($\iota = 90^\circ$) configuration}
        \label{fig:chi_p_posterior_o5_iota90}
    \end{subfigure} 
    \caption{Posterior probability distributions for the spin precession parameter ($\chi_{\mathrm{p}}$) under the O5 noise scenario ($\mathrm{SNR} \approx 35$). Posteriors are shown as violin plots across mass ratios and eccentricities. The dashed gray curve represents the $\chi_{\mathrm{p}}$ prior. In panel (a), higher SNR causes the $\chi_{\mathrm{p}}$ posterior to deviate from the prior for the asymmetric mass ratios ($q \le 0.5$). In panel (b), louder higher-order mode content leads to more tightly constrained, spurious $\chi_{\mathrm{p}}$ posteriors as unmodeled eccentric features are heavily absorbed into the precessing degrees of freedom.}
    \label{fig:chi_p_posterior_o5}
\end{figure}

It is worth noting that across all the injection configurations, the optimal \ac{snr} recovered by the quasi-spherical \ac{pe} pipeline closely tracks the injected $\mathrm{SNR}$. While collecting evidence, the sampler effectively captures all of the signal power by absorbing the unmodelled eccentric amplitude and phase morphologies into the parameter space of the \texttt{IMRPhenomXPHM} approximant, specifically its precessing-spin ($\chi_p$) degrees of freedom, inclination ($\iota$), total mass ($M_{\text{tot}}$), and luminosity distance $d_{\text{L}}$. {\it Therefore, the total \ac{snr} recovered by the \ac{pe} pipeline cannot serve as a diagnostic for missing physics in the case of such short-duration signals originating from high-mass binary mergers.}

To rigorously verify whether the missing eccentric physics leaves a detectable signature in the strain data, we additionally perform a residual test. In this test, we subtract the maximum-likelihood quasi-circular precessing waveform from the whitened detector strain to check whether the remaining data within the signal window is consistent with background noise. Specifically, we compare both the integrated residual energy $E_{\text{res}}$ and the amplitude distribution of the time-domain residual to check for any excess power and departures from Gaussianity using the Kolmogorov--Smirnov (KS) test.

\begin{widetext}
\begin{figure*}[ht!]
    \centering
    \includegraphics[width=0.9\textwidth]{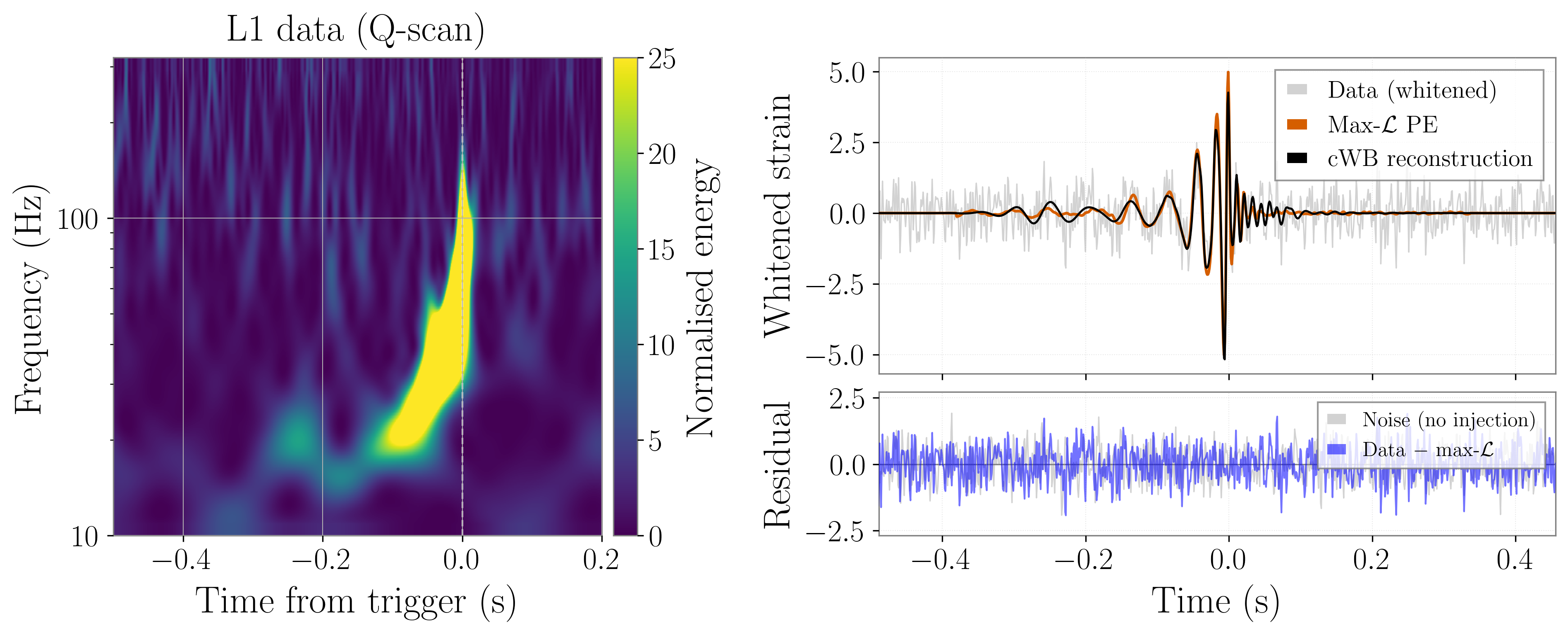}
    \caption{Time-frequency and time-domain residual comparison in L1 for the edge-on ($\iota = 90^\circ$) eccentric injection ($q=0.5$, $e=0.19$) in simulated O5 Gaussian noise. \textit{Left:} Q-transform spectrogram of the whitened strain. \textit{Top right:} Whitened detector strain (grey) overlaid with the quasi-circular precessing $\text{max-}\mathcal{L}$ PE waveform (orange) and the unmodeled cWB reconstruction (black). \textit{Bottom right:} Residual $r(t) = d(t) - h_{\text{max-}\mathcal{L}}(t)$ (blue) compared against a representative off-source noise realization of equal duration (grey).}
    \label{fig:single_event_residual_2532}
\end{figure*}
\end{widetext}

Figure~\ref{fig:single_event_residual_2532} shows the time-frequency spectrogram, waveform comparison, and time-domain residual $r(t) = d(t) - h_{\text{max-}\mathcal{L}}(t)$ in L1 for the edge-on ($\iota = 90^\circ$) eccentric injection (with $q=0.5$, $e=0.19$) under the O5 sensitivity scenario. For this case, residual energy within the signal support window ($378.2$) lies below the off-source noise median ($417.4$, with $P(E_{\text{noise}} \ge E_{\text{res}}) = 0.875$). An empirical KS test calibrated against off-source noise chunks confirms consistency with Gaussian noise ($D_{\text{KS}} = 0.034$, $p_{\mathrm{emp}} = 0.375$). Thus, the precessing model is able to absorb the eccentric features into biased source parameters rather than leaving a statistically significant residual. We verified that this is the case for all the simulations carried out in this study.

\subsubsection{Implications for Astrophysical Inference}

Overall, the results of the Bayesian inference study indicate that identifying orbital eccentricity in signals from \ac{imbh} systems is a pressing challenge for \ac{pe} in realistic detector noise. It is worth noting that a similar Bayesian inference study with different eccentric NR injections was performed by \citet{Xu:2022zza}. \citet{Xu:2022zza} observed that they largely recovered the $\chi_{\text{p}}$ priors in the case of zero-noise injections, but the addition of noise resulted in posteriors with constrained precession for a subset of the noise realizations they considered, and they therefore concluded that the addition of noise makes it possible to reproduce evidence of highly precessing spins in the signal. While we do not perform zero-noise injections in our work, we observe that in the presence of noise, departures from the $\chi_{\text{p}}$ prior become more evident with louder higher-order-mode content in the signal, as can be seen when comparing the face-on and edge-on configurations of the O4 noise scenario (Figures~\ref{fig:chi_p_posterior_o4_iota0} and ~\ref{fig:chi_p_posterior_o4_iota90}).
Therefore, the severity of this problem escalates with both higher-order-mode content and next-generation detector sensitivities. 

{\it Misidentifying orbital eccentricity as spin precession carries severe consequences for astrophysical inference.} Because non-zero $\chi_{\mathrm{p}}$ and orbital eccentricity serve as distinct identifiers for different binary formation channels, falsely inferring precession risks misclassifying dynamical mergers. This misclassification can directly bias population-level inferences of black hole spin-tilt distributions and merger-rate densities for different binary formation channels. It is therefore imperative to investigate whether current model-comparison metrics are sufficient to disentangle these two effects. 

To examine how this degeneracy behaves for candidates with even higher total mass, we repeated the analysis using the same simulation set scaled to GW231123-like masses. Due to the higher masses, the in-band eccentricity ($e_{\mathrm{10Hz}}$) is shifted to lower values, resulting in lower biases in the recovered $\chi_{\text{p}}$ posteriors. A brief discussion regarding this is presented in Appendix~\ref{sec:appendix_231123}.


\subsection{Mass, Distance and Inclination}
\label{sec:pe_m_dl_results}

\subsubsection{O4 Scenario}
\label{sec:pe_m_dl_results_o4}

In the O4 scenario, the recovery of mass, distance, and inclination exhibits biases dependent on the orbital eccentricity and inclination. 
\begin{itemize}
    \item \textbf{Inclination $\iota$ and distance $D_{\mathrm{L}}$:} For face-on injections ($\iota = 0^\circ$), the inclination angle is generally poorly constrained with bimodal posteriors, resulting in poor recovery of distance due to well-known distance--inclination degeneracies in waveforms dominated by the $(\ell, m) = (2, \pm 2)$ mode. 

    \item \textbf{Masses ($M_{\mathrm{tot}}$ and $q$):} For face-on systems ($\iota = 0^\circ$), the total mass is recovered within the 90\% credible interval for the low and moderate eccentricities ($e \lesssim 0.20$). However, at high eccentricity, ($e \sim 0.30$), the quasi-circular PE exhibits biases in total mass recovery to compensate for the unmodeled eccentric phase evolution and amplitude). The mass ratio is not well-recovered as a result of the suppression of higher-order-mode content in the face-on ($\iota = 0^\circ$) configuration. In contrast, for edge-on injections ($\iota = 90^\circ$), the presence of higher-order modes introduces stricter morphological constraints that break this mass--eccentricity degeneracy, resulting in total mass recovery within the 90\% credible interval even for the high eccentricity ($e \sim 0.30$) cases. The injected mass ratio lies within the 90\% credible interval of the recovered posterior for all cases. 
\end{itemize}

\subsubsection{O5 Scenario}
\label{sec:pe_m_dl_results_o5}

For the O5 scenario, the parameter recovery exhibits clear biases that now also include biases due to systematics and not just statistical uncertainty due to the higher \ac{snr}.

\begin{itemize}
    \item \textbf{Inclination $\iota$:} Unlike in the O4 regime where we recovered bimodal posteriors for inclination angle for the face-on injections ($\iota=0^\circ$), the inclination angle is better constrained for the O5 scenario and we do not see bimodal posteriors. For the edge-on injections ($\iota=90^\circ$), the inclination is recovered within the 90\% credible interval for all our injections.
    
    \item \textbf{Masses ($M_{\mathrm{tot}}$ and $q$):} Posteriors for all parameters are much sharper due to the higher \acp{snr}. For both face-on and edge-on cases, $M_{\text{tot}}$ is recovered within or is marginally outside the 90\% credible interval for the low eccentricity cases. However, for the moderate and high eccentricities, the injected $M_{\text{tot}}$ lie significantly outside the 90\% credible interval, with highest biases in the $q=0.33$ case. Similar to the O4 scenario for the face-on configuration, the mass ratio is not well-recovered as a result of the suppression of higher-order-mode content in the face-on ($\iota = 0^\circ$) configuration. In contrast, for the edge-on configurations, the injected mass ratio lies within the 90\% credible interval of the recovered posterior for all cases.
\end{itemize}

\subsection{\ac{cwb} Waveform Consistency}
\label{sec:cwb_reconstruction_results}

Following the parameter estimation results described in section \ref{sec:pe_spin_results}, we quantify the consistency between the precessing waveforms used in \ac{pe} and the unmodeled signal reconstruction using the framework described in section \ref{sec:methods_cwb_waveform_consistency}.

We use $\eta_{0.05}$ to assess the same. Firstly, we use the control injection (section~\ref{sec:results_control_injection_cwb}) to set the baseline.
We note that amongst all the cases, the largest value of the overlap tail fraction for the control injection is for the O5 noise, $\iota=90^\circ$ and is $\eta_{0.05}=0.25$ (See Figure \ref{fig:overlaps_violin_control_o5}). We take a conservative approach and therefore regard a signal with $\eta_{0.05} \leq 0.25$ as consistent with the \ac{pe} waveform, whereas values approaching unity indicate that most of the on-source distribution lies below what is typical for a faithful \ac{pe} template. For reference, identical distributions would give $\eta_{0.05}=0.05$. Please note that this is an arbitrary but reasonable choice within the domain of the study.

\subsubsection{O4 Scenario}
\label{sec:cwb_reconstruction_results_o4}

\begin{figure}[ht!]
\centering
\captionsetup[subfigure]{skip=5pt}  
    \begin{subfigure}{0.49\textwidth}
        \centering
        \includegraphics[width=\linewidth]{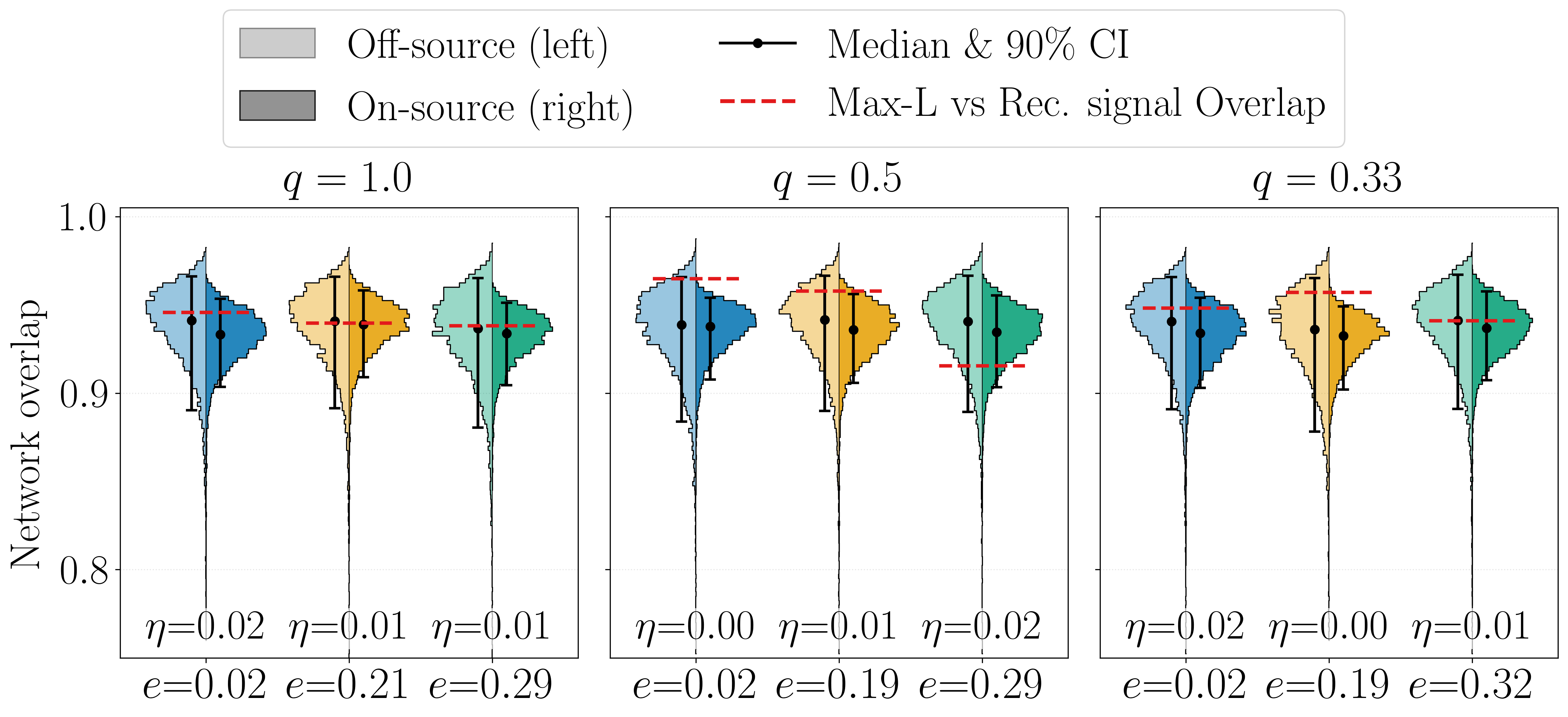}
        \caption{Face-on ($\iota = 0^\circ$) configuration}
        \label{fig:overlaps_o4_iota0}
    \end{subfigure}
    \vspace{0.1cm}
    \begin{subfigure}{0.49\textwidth}
        \centering
        \includegraphics[width=\linewidth]{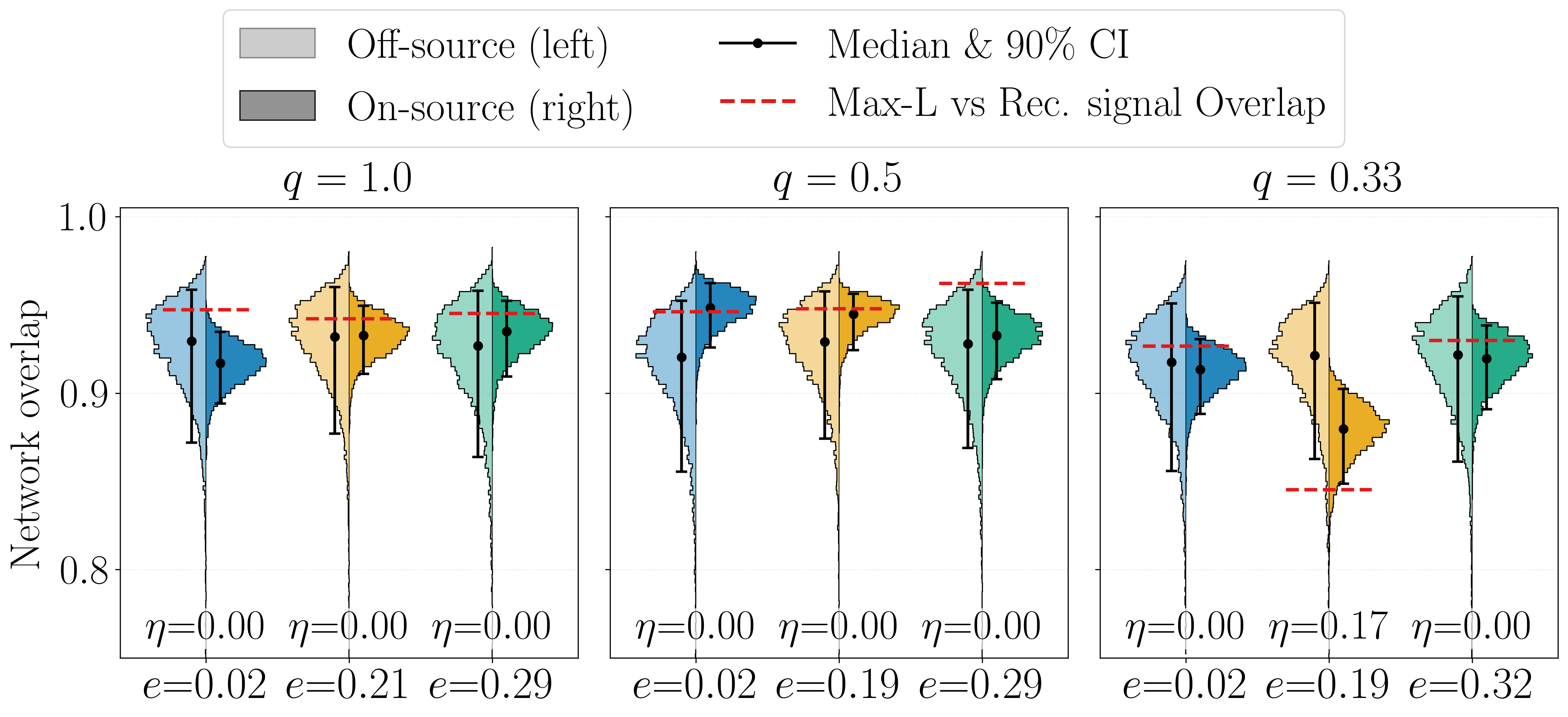}
        \caption{Edge-on ($\iota = 90^\circ$) configuration}
        \label{fig:overlaps_o4_iota90}
    \end{subfigure} 
    \caption{Off-source and on-source overlap distributions for the O4 noise scenario are shown with face-on ($\iota=0^\circ$) configuration in panel (a) and edge-on configuration ($\iota=90^\circ$) in panel (b). Overlap tail fraction, $\eta_{0.05}$ values are quoted in the plots for each case. For both face-on and edge-on configurations, the off-source and on-source overlap distributions are well aligned with low values of $\eta_{0.05}$, with the exception of the $q=0.33, e=0.19$ simulation, where $\eta_{0.05}=0.17$. $\eta_{0.05}<0.25$ for all cases, with the $\eta_{0.05}=0.25$ baseline set by the control injection.}
    \label{fig:overlaps_o4}
\end{figure}

Figure \ref{fig:overlaps_o4} compares the on-source overlap distribution $P(\mathcal{O}_{\text{on}})$ against the off-source null distribution $P(\mathcal{O}_{\text{off}})$ across all eccentricities, mass ratios, and the two inclinations for the O4 noise regime. For both face-on ($\iota = 0^\circ$) and edge-on ($\iota = 90^\circ$) configurations, the off-source and on-source distributions largely overlap across all mass ratios and eccentricities. The only exception occurs for the $q=0.33$, $e=0.19$ case, where mild separation in the histograms is observed, corresponding to an overlap tail fraction $\eta_{0.05} = 0.17$. This demonstrates that at these \acp{snr} ($\mathrm{SNR}=14.5$), the \ac{cwb} waveform consistency test does not possess sufficient resolving capabilities to break the eccentricity--precession degeneracy.

\subsubsection{O5 Scenario}
\label{sec:cwb_reconstruction_results_o5}

\begin{figure}[ht!]
\centering
\captionsetup[subfigure]{skip=5pt}  
    \begin{subfigure}{0.49\textwidth}
        \centering
        \includegraphics[width=\linewidth]{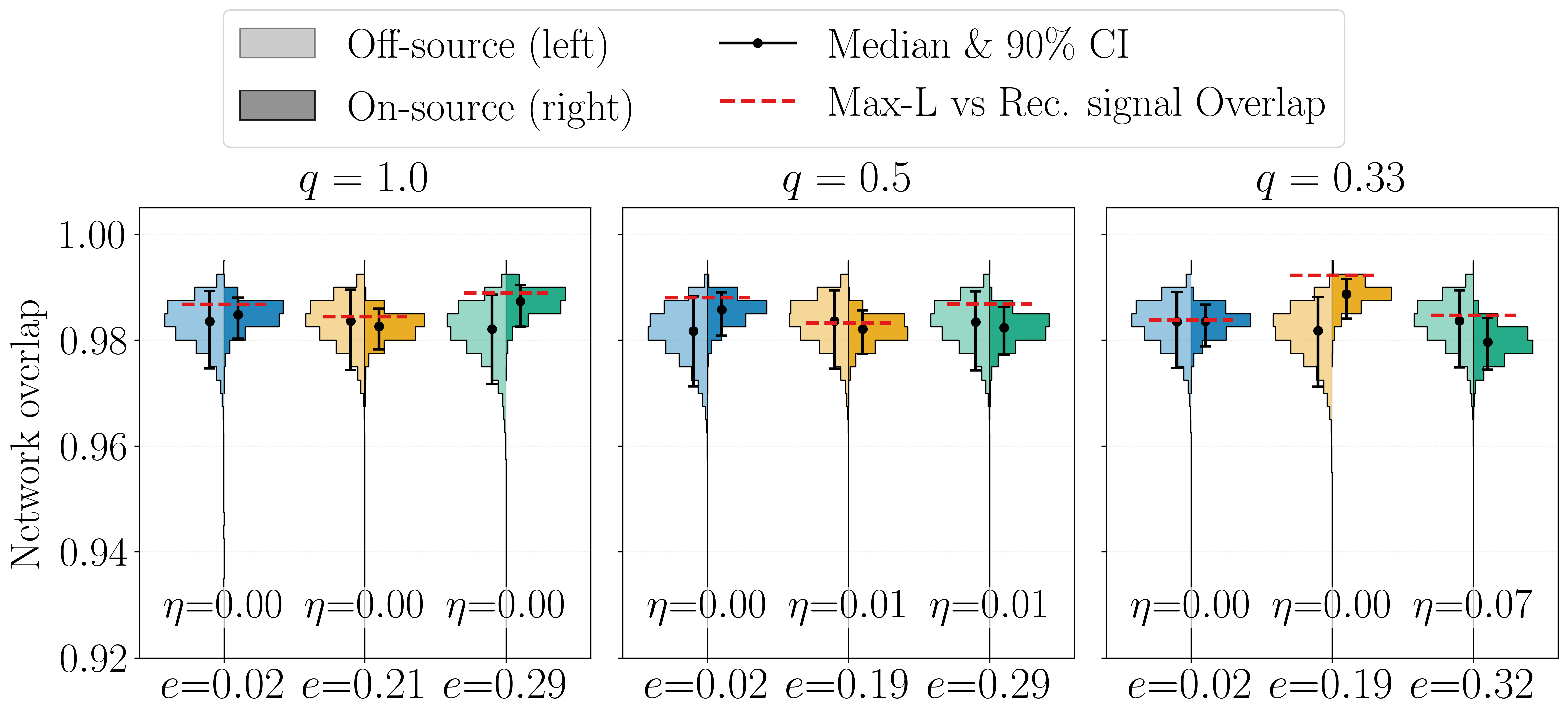}
        \caption{Face-on ($\iota = 0^\circ$) configuration}
        \label{fig:overlaps_o5_iota0}
    \end{subfigure}
    \vspace{0.1cm}
    \begin{subfigure}{0.49\textwidth}
        \centering
        \includegraphics[width=\linewidth]{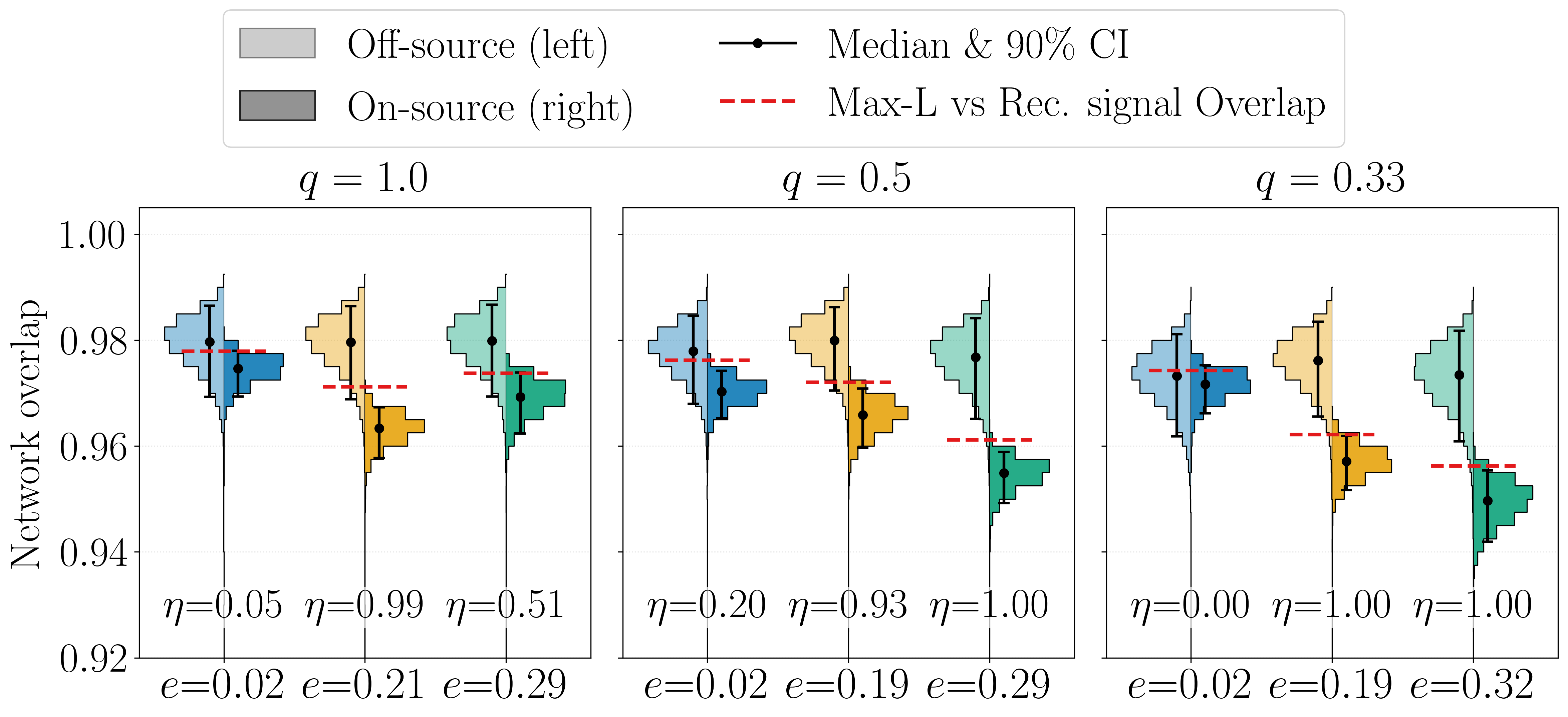}
        \caption{Edge-on ($\iota = 90^\circ$) configuration}
        \label{fig:overlaps_o5_iota90}
    \end{subfigure} 
    \caption{Off-source and on-source overlap distributions for the O5 noise scenario are shown with face-on ($\iota=0^\circ$) configuration in panel (a) and edge-on configuration ($\iota=90^\circ$) in panel (b). Overlap tail fraction, $\eta_{0.05}$ values are quoted in the plots for each case. For the face-on configuration, the off-source and on-source overlap distributions are well aligned, with $\eta_{0.05}<0.25$ for all cases, with the $\eta_{0.05}=0.25$ baseline set by the control injection. However, for the edge-on configuration, we obtain $\eta_{0.05}>0.25$ for all moderate and high eccentricity cases, thereby indicating that the test is able to break the eccentricity--precession degeneracy when high-order-mode content is prominent.}
    \label{fig:overlaps_o5}
\end{figure}

Figure~\ref{fig:overlaps_o5} compares the on-source overlap distribution $P(\mathcal{O}_{\text{on}})$ against the off-source null distribution $P(\mathcal{O}_{\text{off}})$ across all eccentricities, mass ratios, and the two inclinations for the O5 noise regime. In the absence of strong higher-order mode content ($\iota = 0^\circ$), the off-source and on-source distributions largely overlap for all cases across mass ratios and eccentricities (Figure~\ref{fig:overlaps_o5_iota0}). In contrast, for the edge-on systems ($\iota = 90^\circ$), while the two distributions are aligned at low eccentricities ($e \approx 0.02$), they distinctly separate for moderate-to-high eccentricities ($e \ge 0.20$). Since the distributions become sharper owing to the higher SNR, the consistency test is able to resolve subtle mismatches between the quasi-circular PE templates and the unmodeled \ac{cwb} reconstructions. Furthermore, the tail fraction statistic, $\eta_{0.05}$ approaches unity for moderate and high eccentricity cases in the edge-on case, and is effectively able to flag the mismatch between the true signal and \ac{pe} template. {\it These results clearly demonstrate that in the presence of louder higher-order mode content, the proposed consistency test based on the overlap distributions reliably detects the lack of physics information in the waveforms used in PE and breaks the eccentricity--precession degeneracy.} 

\section{Conclusions}
\label{sec:conclusions}


Astrophysical parameter degeneracy is a persistent challenge for short-duration signals, particularly \ac{gw} signals from massive \ac{bbh} observed by the advanced ground-based detector networks. A prime example of this is the first confident \ac{imbh} event, GW190521, whose short-duration, merger-dominated morphology led to a plethora of interpretations for its source parameters. Using GW190521-like signals as a case study, we attempt to break the parameter degeneracy between eccentricity and spin precession using concrete examples across observational scenarios.  

Specifically, we investigate whether quasi-spherical waveform approximants such as \texttt{IMRPhenomXPHM} can mask eccentricity as spin-precession during \ac{pe} of GW190521-like signals in realistic noise scenarios. We then test whether a model-agnostic waveform consistency test based on \ac{cwb} reconstruction can unmask eccentricity by flagging the mismatches between the \ac{pe} template and the reconstructed signal. Using nine eccentric \ac{nr} injections spanning mass ratios and eccentricities, we analysed the results for edge-on and face-on configurations in O4 and O5 noise scenarios.

For \ac{pe} of eccentric injections that are non-spinning, recoveries with \texttt{IMRPhenomXPHM} can return $\chi_{\text{p}}$ posteriors that depart from the prior.
The $\chi_{\text{p}}$ posteriors in O4a noise are largely prior-like for the face-on case ($\iota=0^\circ$) and deviate from the prior for the edge-on case ($\iota=90^\circ$). In the O5 scenario ($\text{SNR} \approx35$), $\chi_{\text{p}}$ posteriors start to deviate from the priors for the  face-on configuration ($\iota=0^\circ$) as well. For edge-on ($\iota=90^\circ$) systems, the $\chi_{\text{p}}$ posteriors deviate from the priors and are significantly more constrained. The posteriors are comparable in sharpness to the control (non-eccentric, precessing) injection. Moreover, standard time-domain residual tests performed by subtracting the maximum-likelihood quasi-circular precessing  \ac{pe} waveform from the strain yield residuals consistent with Gaussian noise across all simulations, showing that the precessing model absorbs the eccentric features into biased parameters rather than leaving excess power.
For signals from \ac{imbh} systems, the $\chi_{\text{p}}$ posterior can therefore be constrained at high values even in the absence of \ac{bh} spins, and an independent check is required to ascertain if this is genuine. 


The \ac{cwb} waveform consistency test serves to perform such a check. In the O4a noise scenario, the off-source and on-source distributions largely align, with low overlap tail fraction statistic, $\eta_{0.05}$ values. 
At this \ac{snr}, the waveform consistency test is therefore not able to resolve mismatches between the true eccentric signal and quasi-circular precessing \ac{pe} templates. 
In O5 noise, the distributions are significantly sharper owing to the higher \ac{snr}. For face-on ($\iota = 0^\circ$) systems the on-source and off-source distributions again largely coincide, while for the edge-on ($\iota = 90^\circ$) systems the on-source distributions shift to lower overlaps for $e\gtrsim0.2$, with $\eta_{0.05}\geq0.93$ in five of the six cases. As expected, $\eta_{0.05}<0.25$ is satisfied by all the low-eccentricity signals. 

In conclusion, we find that the waveform consistency test is effective in a specific regime: at high \acp{snr} and when higher-order-mode content is prominent. The test reliably identifies inconsistencies between a reconstructed eccentric non-spinning signal and quasi-circular precessing \ac{pe} templates, thereby breaking the eccentricity--precession degeneracy. This definitely provides a way forward especially in the future observing runs with improved detector sensitivities as is shown in the study with projected O5 noise. However, the test does not give encouraging results and shows its limitations in breaking the degeneracy at lower \acp{snr}, where the overlap distributions are too broad to resolve the mismatch, nor for face-on systems at O5 \acp{snr}, where the on-source and off-source distributions largely coincide.  


Comparing the results of the analyses on GW190521- and GW231123-like (Appendix~\ref{sec:appendix_231123}) simulations demonstrates that eccentricity--precessing degeneracy is governed directly by the residual in-band eccentricity ($e_{10\mathrm{Hz}}$) when the signal enters the detectors' sensitivity band. For identical eccentric \ac{nr} simulations, scaling to higher total masses shifts the $10\,\mathrm{Hz}$ frequency point closer to merger, lowering $e_{10\mathrm{Hz}}$ and suppressing the eccentric features that otherwise bias $\chi_{\text{p}}$. Probing effects comparable to those found for the GW190521-like injections with moderate to high \textit{in-band eccentricity} would therefore require \ac{nr} simulations with higher eccentricity at the \ac{nr} reference frequency. We leave this to future work.

These findings establish a concrete strategy to address the parameter degeneracy due to missing physics in the waveform models used in PE studies. The cWB waveform-consistency-based metric for loud signals where the degeneracy is expected to be large, namely high mass, high mass ratios and high (\textit{in-band}) eccentric edge-on systems is capable to break this degeneracy. Though, our study appears as a case study, specific and restricted to GW190521-like systems, this regime captures the parameter space where template misspecification is most acute and relevant for the LIGO--Virgo--KAGRA detector network.

Some caveats of this work motivate future directions to improve upon this method. Most notably, our baseline threshold for the overlap tail fraction, $\eta_{0.05}<0.25$ in this work is determined using a single control injection. Future work would include generating a large number of control injections with varying source parameters and across multiple noise realizations which would enable the construction of an empirical null distribution of $\eta_{0.05}$ values. This would allow for a more statistically robust method for ranking the significance of a true eccentric candidate's $\eta_{0.05}$ statistic. Additionally, expanding the parameter space beyond extreme inclinations ($\iota=0^\circ$ and $\iota=90^\circ$) and non-spinning eccentric injections to include intermediate inclinations and eccentric simulations with spins (both aligned and precessing) will improve our understanding of how these effects can jointly impact the validity of this test.

\begin{acknowledgments}
The authors gratefully acknowledge Tanmaya Mishra, Rimo Das, and Edoardo Milotti for valuable discussions. The authors also thank Marek Szczepanczyk and Juan Calderon Bustillo for their insightful comments during the \ac{lvk} internal circulation.
The authors additionally thank the SXS collaboration for maintaining their public waveform database. 
This material is based upon work supported by NSF's LIGO Laboratory which is a major facility fully funded by the National Science Foundation. The authors acknowledge the computational resources that aided the completion of this project, provided by LIGO-Laboratory and supported by the NSF Grants No..PHY-0757058 and No.PHY-0823459. 
SB acknowledges support from IIT-Bombay. JF acknowledges support from the Prime Minister’s Research Fund.
AP acknowledges support from ANRF/ARGM/2025/001361/TS.

\end{acknowledgments}

\appendix

\section{NR Injection Set}
\label{sec:appendix_nr_set}

Table~\ref{tab:nr_waveforms} lists NR simulations used in this study with their respective binary parameters. The eccentricities quoted are computed using the \texttt{gw-eccentricity} package at $10 \text{ Hz}$, assuming a source-frame total mass of $M_{\text{tot}} = 150\,M_{\odot}$. While the initial eccentricities of these simulations are high (up to $e \approx 0.8$), due to the circularization of the binary's orbit via \ac{gw} emission, the eccentricity efficiently decays by the time the signal from such a massive system reaches $10 \text{ Hz}$. For each mass ratio, we chose
simulations with low ($e \approx 0.02$), intermediate ($e \approx 0.2$), and high ($e \approx 0.3$) eccentricity at $10\,\mathrm{Hz}$, so that the eccentricities are comparable across mass ratios. Each simulation is injected at two inclinations, $\iota = 0^\circ$ (face-on) and $\iota = 90^\circ$ (edge-on) to understand the contribution of higher-order-mode content. Additionally, we include SXS:BBH:0391 as a quasi-circular, precessing control injection.

\begin{table}[hbt!]
\setlength{\tabcolsep}{7pt} 
\centering

\begin{tabular}{lcccc}
\hline
SXS ID & $q$ & $e$ & $\chi_p$ & $\iota$ \\
\hline
\\[-5pt] 
\multicolumn{5}{c}{Eccentric injections} \\[2pt]
\hline
SXS:BBH:2518 & 1.0  & 0.02 & 0.0 & $0^\circ, 90^\circ$ \\
SXS:BBH:3972 & 1.0  & 0.21 & 0.0 & $0^\circ, 90^\circ$ \\
SXS:BBH:2524 & 1.0  & 0.29 & 0.0 & $0^\circ, 90^\circ$ \\
SXS:BBH:2529 & 0.5  & 0.02 & 0.0 & $0^\circ, 90^\circ$ \\
SXS:BBH:2532 & 0.5  & 0.19 & 0.0 & $0^\circ, 90^\circ$ \\
SXS:BBH:2534 & 0.5  & 0.29 & 0.0 & $0^\circ, 90^\circ$ \\
SXS:BBH:2536 & 0.33 & 0.02 & 0.0 & $0^\circ, 90^\circ$ \\
SXS:BBH:2543 & 0.33 & 0.19 & 0.0 & $0^\circ, 90^\circ$ \\
SXS:BBH:2544 & 0.33 & 0.32 & 0.0 & $0^\circ, 90^\circ$ \\
\hline
\\[-5pt] 
\multicolumn{5}{c}{Control injection} \\[2pt]
\hline
SXS:BBH:0391 & 0.5  & 0.0    & 0.34 & $0^\circ, 90^\circ$ \\
\hline
\end{tabular}

\caption{Numerical-relativity waveforms used for the injections. The mass ratio is defined as the ratio between the lighter and heavier mass, $q = m_2/m_1$, $e$ is the orbital eccentricity computed at $10 \text{ Hz}$ in the source frame, and $\chi_p$ is the effective precession spin. Each simulation is injected at two inclinations, $\iota = 0^\circ$ (face-on) and $\iota = 90^\circ$ (edge-on). The last row is the quasi-spherical control injection.}
\label{tab:nr_waveforms}
\end{table}


\section{GW231123}
\label{sec:appendix_231123}

Detected during O4a, GW231123 surpassed GW190521 as the most massive \ac{bbh} merger detected so far by the advanced ground-based \ac{gw} detectors, with a total source-frame mass of $190-265 \, M_{\odot}$, component masses of $137^{+22}_{-17}\,M_\odot$ and $103^{+20}_{-52}\,M_\odot$ and high spin magnitudes for both black holes estimated to be $0.90^{+0.10}_{-0.19}$ and $0.80^{+0.20}_{-0.52}$ \cite{LIGOScientific:2025rsn}.
To examine the effect of higher total mass (and equivalently, shorter duration of the signal), we repeat the end-to-end analysis using GW231123-like injections. For the O4 scenario, the luminosity distance $d_L$ is scaled for all nine configurations such that the network SNR is a  GW231123-like signal-to-noise ratio of $\text{SNR} = 22.5$. Signals are injected at the same GPS time as the GW190521-like injections (GPS time of 1379824500).

\begin{figure}[h]
    \centering  \includegraphics[width=0.44\textwidth]{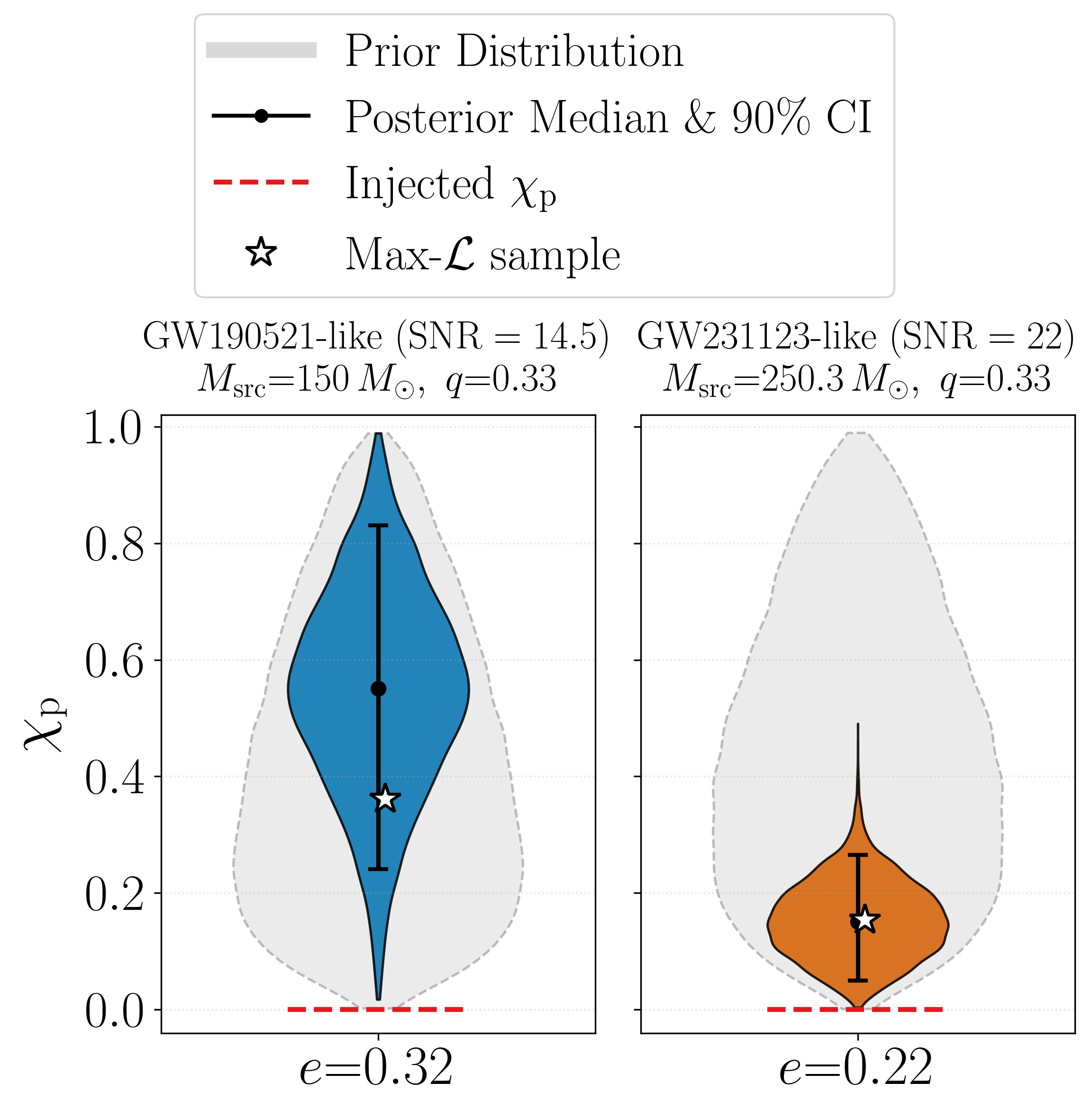}
    \caption{Recovered $\chi_{\text{p}}$ posteriors for the same \ac{nr} simulation (\texttt{SXS:BBH:2544}) scaled to GW190521-like masses (left) and GW231123-like masses (right). $e_{\text{10Hz}}$ computed using the \texttt{gw-eccentricity} package for each case is indicated at the bottom. The effect of the lower in-band eccentricity translates to a lower constraint of $\chi_{\text{p}}$, with the posteriors being sharper for the GW231123-like scenario owing to its higher \ac{snr}.}
    \label{fig:231123_190521_chip_2544}
\end{figure}

Unlike the GW190521-like case, the \ac{pe} recoveries for the GW231123-like injections return uninformative $\chi_{\text{p}}$ posteriors or $\chi_{\text{p}}$ posteriors with significantly lower median. This behaviour is expected since the in-band eccentricities ($e_{\text{10Hz}}$) of these signals is lower. This is due to the fact that for the higher total mass of GW231123, a frequency of $10\text{ Hz}$ is reached later in the inspiral. By this point, radiation reaction has efficiently circularised the orbit such that the in-band eccentricities computed using the \texttt{gw-eccentricity} package \cite{Shaikh:2023ypz} is $e_{10 \text{Hz}} \approx 0.2$ for the highest eccentricity in the simulation set that we use. 
Figure~\ref{fig:231123_190521_chip_2544} shows the recovered $\chi_{\text{p}}$ posteriors for the same \ac{nr} simulation (\texttt{SXS:BBH:2544}) scaled to GW190521-like masses (left) and GW231123-like masses (right). The lower in-band eccentricity results in the recovered $\chi_{\text{p}}$ posterior being constrained to lower values, with being sharper dostributions for the GW231123-like scenario owing to its higher \ac{snr}.


\section{PE Systematics: Waveform Systematics and Noise Realization Anomalies}
\label{sec:appendix_pe_systematics_noise_anomalies}

\subsection{Waveform Systematics in \texttt{IMRPhenomXPHM}}
\label{sec:appendix_pe_systematics}

For our control injection ($q=0.5, \chi_{\text{p}}=0.34, e=0$) in the O5 noise scenario, the injected value of $\chi_{\text{p}}$ lies outside the 90 \% confidence interval of the recovered posterior for the edge-on ($\iota = 90^\circ$) configuration, as can be seen in Figure \ref{fig:pe_control_chi_p_o5}. Since the control injection is generated from a quasi-circular precessing \ac{nr} simulation, this bias cannot be attributed to eccentricity, or any other unmodeled physics. This points to possible waveform systematics in the \texttt{IMRPhenomXPHM} approximant. 

To verify this, we computed the (O5) noise-weighted match (overlap) between the \ac{nr} simulation and an  \texttt{IMRPhenomXPHM} waveform generated using identical initial parameters and reference frequency as the \ac{nr} simulation. For $\iota = 90^\circ$, we obtain a match of $97\%$, with high mismatches originating from the $(\ell, m) = (2, \pm 1)$ and $(\ell, m) = (3, \pm 2)$ modes. At edge-on orientations, the $(\ell, m) = (2, \pm 2)$ quadrupole mode is relatively suppressed, which can expose waveform systematics in the subdominant modes. For high \acp{snr} like what we have in the O5 noise scenario, this mismatch exceeds systematic-error thresholds and can result in biases in the inferred posteriors.

\subsection{Noise-Realisation Anomalies}
\label{sec:appendix_noise_anomalies}
In addition to waveform systematics, \ac{pe} of moderate \ac{snr} signals in real detector noise can be susceptible to anomalies due to non-stationary noise fluctuations. Within our O4a real-noise injection suite ($\mathrm{SNR} = 14.5$), one configuration initially led to spurious \ac{pe} results characterized by a peak at a suspiciously low mass ratio value.

Because \ac{gw} signals from \ac{imbh} systems are short in duration and span few cycles in the ground-based detectors' sensitive band, non-Gaussian features in real data may correlate with longer-duration, lower-mass templates, resulting in a local maximum in the likelihood surface, which can lead to spurious \ac{pe} posteriors. To verify that this was indeed the case, we shifted the injection time for this simulation by $1 \text{s}$, while keeping all other parameters fixed. Repeating \ac{pe} analysis on this shifted injection fixed the issue with mass posteriors as expected.


\newpage
\bibliography{references.bib} 


\end{document}

%% file: flowchart.tex
\begin{tikzpicture}[
    node distance=0pt,
    >={Stealth[length=2.4mm]},
    box/.style={draw, rounded corners=2pt, align=center, text width=4.9cm,
              minimum height=1.1cm, inner sep=4pt, font=\small, thick},
    gen/.style={box, fill=gray!12},
    bil/.style={box, draw=cbBlue, fill=cbBlue!12},
    cwb/.style={box, draw=cbOrange!85!black, fill=cbOrange!18},
    dist/.style={box, draw=cbGreen!70!black, fill=cbGreen!14},
    cmp/.style={box, draw=cbPurple!80!black, fill=cbPurple!20},
    arr/.style={->, thick},
]
\definecolor{cbPurple}{HTML}{CC79A7}
\node[gen] (h) at (3.2,0)     {Numerical-relativity waveform $h$\\
                              (scaled to GW190521-like source-frame mass)};
\node[gen] (s) at (3.2,-2.2) {Inject into detector noise $n$\\
                              $d = n + s$\\
                              {\footnotesize (O4a real noise / O5 Gaussian noise)}};
\draw[arr] (h) -- (s);

\node[bil] (bil) at (0,-4.8)   {Bayesian inference on $d$\\ ({\tt Bilby}, using quasi-circular model \tt IMRPhenomXPHM)};
\node[cwb] (cwb) at (6.4,-5.95) {Signal reconstruction\\ of $s$ with cWB, $\hat h_s$};

\draw[thick] (s.south) -- ++(0,-0.5) coordinate (fork);
\draw[arr] (fork) -| (bil.north);
\draw[arr] (fork) -| (cwb.north);

\node[bil] (post) at (0,-7.1)   {Posterior samples $\{\theta_i\}_{i=1}^{N}$\\
                                  {\footnotesize (incl.\ max-$\mathcal{L}$ sample)}};
\draw[arr] (bil) -- (post);

\node[cwb]  (inj) at (0,-9.7) {For each $\theta_i$: inject $h(\theta_i)$ into\\
                               off-source noise and reconstruct\\ with cWB $\rightarrow \hat h_i$};
\draw[arr] (post) -- (inj);

\node[dist] (off) at (0,-12.3)   {\textbf{Off-source} (null) distribution\\
                                  $\mathcal{O}_{\rm off}^{(i)} = \mathcal{O}\!\left[h(\theta_i),\hat h_i\right]$};
\draw[arr] (inj) -- (off);

\node[dist] (on) at (6.4,-12.3) {\textbf{On-source} distribution\\
                                 $\mathcal{O}_{\rm on}^{(i)} = \mathcal{O}\!\left[h(\theta_i),\hat h_s\right]$};
\draw[arr] (cwb) -- (on);
\draw[arr] (post.east) -- ++(0.6,0) |- (on.west);

\node[cmp] (cmp) at (3.2,-14.6) {Compare $P(\mathcal{O}_{\rm on})$ with $P(\mathcal{O}_{\rm off})$\\
                                 {\footnotesize (tail fraction $\eta_\alpha$)}};
\draw[arr] (off.south) |- (cmp.west);
\draw[arr] (on.south) |- (cmp.east);

\end{tikzpicture}